\documentstyle[aps,epsf,psfig]{revtex}
\twocolumn
\begin{document}
\twocolumn[\hsize\textwidth\columnwidth\hsize\csname @twocolumnfalse\endcsname

\title{Theory for Metal Hydrides with Switchable Optical Properties}
\author{K. K. Ng$^{1,2}$, F. C. Zhang$^{1,3}$, V. I. Anisimov$^4$,
and T. M. Rice$^5$}
\address{
$^1$Department of Physics, University of Cincinnati,
Cincinnati, Ohio 45221\\ 
$^2$Yukawa Institute for Theoretical Physics, Kyoto University, Kyoto 606-01, 
Japan\\ 
$^3$Institute of Physics, Academia Sinica, Taipei, Taiwan\\
$^4$Institute of Metal Physics, Russian Academy of Sciences, 620219,
Ekaterinburg, GSP-170, Russia\\
$^5$Theoretische Physik, ETH-H\"onggerberg, 8093 Z\"urich, Switzerland\\
{\rm(\today)}}
\maketitle

\begin{abstract}
Recently it has been discovered that lanthanum, yttrium, and other metal
hydride films show dramatic changes in the optical properties at the
metal-insulator transition. Such changes on a high energy scale suggest
the electronic structure is best described by a local model based on 
negatively charged hydrogen (H$^-$) ions. We 
develop a many-body theory for the strong correlation in a H$^-$ ion lattice. 
The metal hydride
is described by a large $U$-limit of an Anderson lattice model. We use 
lanthanum hydride as a prototype of these compounds, 
and find LaH$_3$ is an insulator with a substantial gap consistent with 
experiments. It may be viewed either as a Kondo insulator or a band
insulator due to strong electron correlation. A H  vacancy state in LaH$_3$ is 
found to be highly localized due to the strong bonding between the electron
orbitals of hydrogen and metal atoms. Unlike the impurity states in the usual
semiconductors, there is only weak internal optical transitions within
the vacancy. The metal-insulator transition takes place in a band of
these vacancy states. 

\end{abstract}
\vskip2pc]


\section{Introduction}

Huiberts $et$ $al.$ \cite{Huiberts1} have recently reported dramatic changes 
in the optical properties of lanthanum, yttrium, and other rare earth hydride 
films \cite{Vajda} with changing hydrogen content. This phenomenon has also 
been recently observed  in other metal hydrides, such as Gd-Mg alloy hydrides 
\cite{Van}. By changing the hydrogen gas pressure, or by electrochemical 
means \cite{fcz6} the films can be continuously 
and reversibly switched from a shiny mirror (good metal) to a transparent 
window (insulator) in a fraction of a second. Although many 
metal-insulator transitions are known, this type of switchable optical 
phenomena is 
very unusual, and potentially of considerable technological importance since 
the transition leads to spectacular effects in the $visible$ light region.
 For example, this class of materials is very 
different from the  previously reported charge density wave compound KMoO$_3$.
In that compound, the frequency dependent conductivity may be changed by
replacing some of Mo atoms by W\cite{fcz3}, but the change in
reflectivity is in the infrared  or lower frequency region, while the
metal hydrides show dramatic changes in the visible. 
Since the hydrides are easily tunable, this class of compounds is also 
ideal for basic research to understand the metal-insulator transition in 
general.  Furthermore, since the optical switching is realized at room 
temperature and at normal pressure, and since there appears to be 
considerable scope for shortening the time-scale of the transition through 
chemical or electrochemical means\cite{Griessen}, the phenomenon is very 
attractive for optical devices. 

As Huiberts $et$ $al$. \cite{Huiberts1} reported, the transitions measured in 
the optical transmission and in the electric resistivity appear at the same
hydrogen concentration $x=x_c\simeq 2.80$ for both the YH$_x$ and LaH$_x$. 
The optical gap in the insulating phase is $\sim 1.8$~eV for LaH$_3$ and 
$\sim2.8$~eV for YH$_3$. The transition is clearly of electronic origin in the
lanthanum hydrides where the crystal structure remains face centered 
cubic (fcc) for $2 \leq x \leq 3$. In the yttrium hydrides, there is evidence
that the observed metal-insulator transition is also of electronic origin
\cite{Griessen}.

The electronic structure underlying this behavior was poorly understood $-$ 
indeed the standard local density approximation (LDA) calculations failed to
predict a metal-insulator transition at all. In 1970, Switendick 
\cite{Switendick} used a non-self-consistent approximation to calculate
the electronic structure of the YH$_3$, and found an energy gap of 1.5 eV.
But more sophisticated state-of-the-art self-consistent LDA calculations by 
Dekker $et$ $al.$ \cite{Dekker}, Wang and Chou \cite{Wang} predict that 
LaH$_3$ and YH$_3$ are metals or semimetals.

\begin{figure}[htbp]
\vspace{0.5cm}
\centerline{\psfig{figure=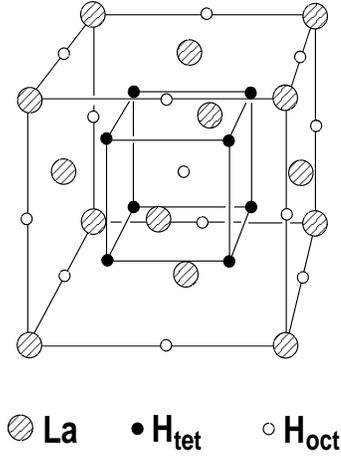,height=6cm}}
\vspace{0.5cm}
\caption{Lattice structure (f.c.c) of LaH$_x$. As $x$ increases from 2 to 3
the H-atom content at octahedral sites increases from empty to full, and a
shiny metal evolves to an insulator. \label{LaH3}}
\end{figure}

The structure changes are especially small in the LaH$_x$ films, which may
be considered as a prototype for this class of compounds. The La atoms always
form a fcc lattice with two H atoms occupying tetrahedrally coordinated
sites. As $x$ changes from 2 to 3, the octahedrally coordinated sites go
from empty to fully occupied and a good metal evolves to a transparent 
insulator. The crystal structure is shown in Fig.~\ref{LaH3}. 
The actual lattice constants of the dihydride and the trihydride are 
10.7051 a.u. (1 a.u.= 0.528~\AA) and
10.5946 a.u. respectively. The slight contraction due to the addition of H atom
in LaH$_3$ indicates strong hybridization between La and H atoms. But the 
small difference in lattice constant is insignificant, and will be neglected 
in our further consideration.

In the metal hydrides, the hydrogen atom attracts one more electron and forms
a negative hydrogen ion H$^-$. The rare earth metal such as La has a formal
valence 3+. In a dihydride such as LaH$_2$, the 1$s$-orbital of the hydrogen 
bands (H bands) are
filled, leaving 1 electron per unit cell in the La conduction bands. This 
gives metallic behavior. In a trihydride such as LaH$_3$, the H bands can hold
all six valence electrons per unit cell. However, in the LDA calculations 
there is an
overlap between the H and La $5d$-orbital bands, leading to a metal or 
semimetal rather
than the observed transparent insulator. The H$^-$ ion is a difficult case
for the LDA, and a careful treatment of the correlation between the two 
electrons is required in order to obtain the bound H$^-$ ion state with 
binding energy of 0.7 eV
(see next section). The poor treatment of correlations is most likely the
reason why LDA is unable to explain the transparent insulating behavior of 
LaH$_3$. It is well known that the LDA underestimates the energy gap in 
semiconductors. This problem seems much more pronounced in rare earth
hydrides: LDA predicts a metallic state for an insulator with substantial gap.

There has been considerable theoretical activity since the original discovery 
\cite{Ng,Eder,Wang,Kelly,Chang}. In an early Letter \cite{Ng}, we examined the
role of electron correlation and proposed a local model to describe the 
electronic structure and the metal-insulator transition in these hydrides. The
purpose of the present paper is to refine our calculations and to
present a more complete theory for these hydrides. Recently, Eder $et$
$al.$ \cite{Eder} have proposed a theory based on the
occupation-dependent hoppings on H sites and examined a
Kondo-like-insulator model for YH$_3$. Their theory has much in common
with ours. Very recently Chang $et$ $al$. \cite{Chang} reported GW method 
calculations for LaH$_3$ and found a gap of $\sim 0.5$~eV. This again indicates
the importance of electron correlation. On the other hand,
Kelly $et$ $al.$ \cite{Kelly} proposed that the insulating nature of
YH$_3$ may be explained within the LDA due to a more complicated
hexagonal structure for Y. Their proposal, however, has difficulty
explaining the insulating nature of many other hydrides, such as
La hydrides, whose crystal structure is a always simple fcc.

In this paper, we examine the importance of electron 
correlation in metal hydrides, and develop a many-body theoretic framework to
study their electronic structure. LaH$_3$ may be viewed as a Kondo insulator
\cite{Ng,Eder} if we start with metallic phase LaH$_2$ and consider every 
additional H atom to a be Kondo impurity. LaH$_3$ may also be viewed as a band
insulator due to the strong electron correlation, which suppresses the 
overlaps between the H and metal bands. When the H concentration is
reduced in the LaH$_3$ structure, the elementary entities are the
H-vacancies. We find highly localized electronic states centered on
these vacancies. The metal-insulator transition takes place in a band
of these vacancy states. The paper is organized as follows. 

In section II, we first review an appropriate wavefunction for a single H$^-$
ion in free space incorporating the electron correlation explicitly. We then
extend the discussion to a lattice of H$^-$ ions, and use a microscopic model
to calculate the electron motion (or the effective hopping integrals) in such
a lattice.

In section III, we propose that the metal hydrides are best 
described by a large $U$-limit
Anderson lattice Hamiltonian, with $U$ the electron Coulomb repulsion on the
H atom. At every H site, electron state is either singly or doubly occupied.
The electron hopping integrals between H sites are described by those 
obtained in our microscopic calculations in section II. Other parameters are
extracted from the LDA calculations. We use the
Gutzwiller method to study this model Hamiltonian for the hydrides within this
method the local constraint on the H sites is replaced by a set of 
renormalization factors.

In section IV we present our results for the electronic structure of LaH$_2$ 
and LaH$_3$. The former is found to be a metal, and the latter an insulator,
in agreement with experiments. 

Section V consists of the study of the localized H vacancy state in LaH$_3$
and the H impurity state in LaH$_2$. The localized vacancy state
is proposed to be
the elementary entity in the hydrides and to be responsible for the 
metal-insulator transition. We will use finite size calculations to verify
the symmetry and localized nature of the vacancy states. In
section VI we use the
calculated electronic band structure to study the optical conductivity and 
density of states. We perform finite size calculations and compare the
results to experimental data.


\section{Electron correlation in negatively charge hydrogen ions} 

In this section we start by reviewing the 
electron correlation in a single negatively charged hydrogen ion
(H$^-$ ion) and the importance of the cooperative motion of the two electrons 
around the proton. We will then describe a local model for studying the 
many-body problem of a lattice of H$^-$ ions. A microscopic model calculation
will be presented to estimate the effective kinetic energy for the electrons
in a H$^-$ ion lattice. 


\subsection{A single H$^-$ ion in free space}

This is a venerable but interesting problem 
\cite{Bethe,Chandrasekhar,Hylleraas,Taylor,Fischer}. It was carefully
studied by astrophysicists back in the 1950's, when the H$^-$ ion was found to
 be of great importance for the opacity of the atmosphere of the sun and 
similar stars. Since hydrogen contains only one proton, the smallest
charge of all nuclei, 
the Coulomb attraction between an electron and its nucleus is 
the weakest and when an additional electron is added to a neutral H
atom, the Coulomb repulsion between the two electrons becomes
crucially important.  
Electron-electron correlation has the largest effect in H atoms compared to
other atoms.  
As found by Chandrasakhar \cite{Chandrasekhar}, Bethe and Salpeter 
\cite{Bethe}, and Hylleraas $et$ $al.$ \cite{Hylleraas}, a careful treatment 
of the correlation between the two electrons is necessary to obtain the correct
binding energy of about 0.7 eV. This shows that the H$^-$ ion is a bound 
state due to a strong electron correlation effect.

\begin{figure}[htbp]
\vspace{0.5cm}
\centerline{\psfig{figure=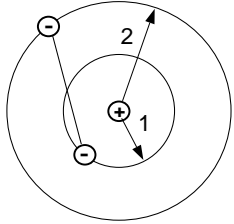,height=3cm}}
\vspace{0.5cm}
\caption{Illustration of Chandrasakhar's wavefunction Eq.~(\protect\ref{psi})
for H$^-$, describing two electrons bound to a proton. The solid line between
the two electrons represents the correlation term in Eq.~(\protect\ref{psi}).
\label{h1}} 
\end{figure}

Variational trial wavefunctions have been applied to the H$^-$ ion for there 
is no exact analytic solution for the ion. Wavefunctions with up to 24 
variation parameters \cite{Hylleraas} were proposed to give the best estimate 
of the binding energy. On the other hand, there is a simple but excellent 
trial wavefunction introduced by Chandrasakhar in the 1940's 
\cite{Chandrasekhar} to describe the electron correlation in H$^-$. 
\begin{equation}
\psi(1,2)= (e^{-a r_1-br_2}+e^{-ar_2-br_1})(1+c|{\bf r}_1-{\bf r}_2|)\chi,
\label{psi}
\end{equation}

\noindent where $({\bf r}_1, {\bf r}_2)$ are the electron coordinates 
with respect to the proton, $\chi$ is the spin singlet spinor,
 and the constants $a$ = 1.075, $b$ = 0.478, $c$ = 0.312 in 
atomic units. As illustrated in Fig.~\ref{h1}, the wavefunction can roughly 
be visualized as an inner electron of radius $\sim 1$, and an 
outer electron of radius $\sim 2$  orbiting around the  proton. The 
electron correlation is 
described by the term $c|{\bf r}_1 -{\bf r}_2|$, which tends to keep 
the two electrons apart. This is similar to the
Laughlin's wavefunctions for the fractional quantum Hall states, where
 electrons in high magnetic fields tend to stay apart and form a quantum 
liquid \cite{Laughlin}. One should notice that the second electron does not 
reside in the same orbital as the first electron although they have opposite 
spins, nor does it occupy the $2s$ orbital which has a very different radial 
wavefunction. In this wavefunction the electrons are in a spin singlet 
state while the spin triplet state is unbound. 
The wavefunction of Eq.~(\ref{psi}) gives a ground state energy of H$^-$ ion 
very close to the best estimate of Hylleraas \cite{Hylleraas} using 24 
variational parameters. Note that the choice of the form of the variational 
wavefunction is very subtle and other simple choices mostly give an unbound 
electron. The standard LDA calculations for a single H$^-$ do 
not give a bound state \cite{Gunnarrson}.

The simple wavefunction of Eq.~(\ref{psi}) will 
be used to develop a many body theory for the H$^-$ ion lattice.
The more refined 24 parameters wavefunction may produce a better result, but 
the calculation of hopping integrals involving such a complicated wavefunction
will be very difficult and time consuming.


\subsection{H$^-$ ions in crystal}\label{Hion}

Since a single H$^-$ ion is a bound state, the ionic picture should  
be valid in certain hydride crystals. A local description is suggested by
the recently observed insulating gap in LaH$_3$ and YH$_3$. This led us to 
examine the effect of correlations on the H$^-$ band width. In this section we
shall focus on H$^-$ ions. The effect of rare earth ions through their 
hybridization with H$^-$ ions is also significant, and will be discussed in 
a later section. The form of the H$^-$ bands in a many body theory is 
determined by the spectrum of the states obtained by removing an electron 
from a lattice of H$^-$ ions. We will use a local model, namely an orthogonal
tight-binding model to describe the hole motion. Then the energy spectrum
is determined  by  the effective 
hopping integrals $t(d)$ of an electron from a H$^-$ ion to a neutral 
H atom at distance $d$. Below we shall estimate $t(d)$ from a microscopic 
model, in which the correlated Chandrasakhar wavefunction Eq.~(\ref{psi}) 
will be employed.

\begin{figure}[htbp]
\vspace{0.5cm}
\centerline{\psfig{figure=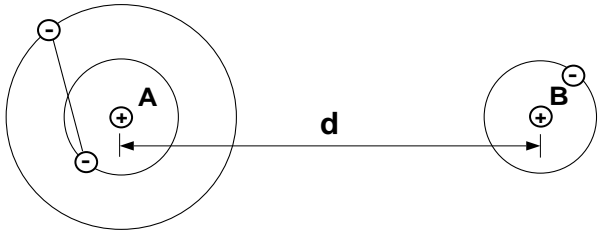,height=3cm}}
\vspace{0.5cm}
\caption{Illustration of H$_2^-$ ion, a system with two protons of 
distance $d$ and three 
electrons.  Shown at left is a H$^-$ ion, described by the wavefunction of 
Eq.~(\protect\ref{psi}), and at right is a neutral H atom in the groundstate.
\label{h2} }
\end{figure}

We consider a single pair consisting of one H$^-$ ion and one H atom separated 
by a distance $d$ 
as illustrated in Fig.~\ref{h2}.
This results in a H$_2^-$ ion, an ion with three electrons moving around two
protons. Previous studies have shown that H$_2^-$ ion is a bound negative ion 
\cite{Taylor} for $d >3$ a.u., the range of interest here (the nearest H-H 
distance separation in LaH$_3$ is 4.58 a.u.). For $d<3$ a.u., H$_2^-$ ion 
becomes  unstable and dissociates into H$_2$ and an unbound, free electron. 
When a H$^-$ ion and H atom are brought together from infinity, the 
ground state manifold of H$_2^-$ splits into odd($-$) and even(+) parity 
states with respect to the center of mass of the two protons. It is
this splitting that determines the effective hopping matrix element $t(d)$.
Let $E_{-}$ and $E_{+}$ be the energies of the odd and even parity states 
respectively, then we have 

\begin{equation}
t(d)=\left(E_{-}-E_{+}\right)/2.
\label{td}
\end{equation} 

\noindent This relation may be understood as follows. Consider an electron 
which hops between two atoms of the same atomic energy with hopping 
integral $t$. Then, the energy of the even parity state is $t$ and the energy 
of the odd parity state is $-t$, the energy difference between the 
two states is $(E_+-E_{-})=2t$. In the H$^-_2$ ion problem, the singlet spin 
gives an additional sign change, as does Eq. (\ref{td}).

The H$_2^-$ ion is a three-body  problem and therefore no exact analytic 
solution is available. Yet a suitable approximation can still give a reasonably
good ground state energy. 
We construct the lowest energy states of different  parities 
using the single site states of H$^-$ ion, Eq.~(\ref{psi}), and the neutral 
H-atom. Let $\phi(\alpha)$ be the hydrogen ground state wavefunction 
(the Bohr atom solution) of the $\alpha$th electron (Fig.~\ref{h2}). Then the 
three electron states of the H$^-_2$ with odd($-$) and 
even(+) parities are given by: 

\begin{equation}
\Psi_{\pm} = \Phi_{i,j} \pm \Phi_{j,i},
\end{equation} 

\noindent where 

\begin{equation}
\Phi_{i,j} =  A [\psi_i(1,2) \phi_j(3)]. 
\end{equation}

\noindent $A$ is the antisymmetric operator to assure the anti-symmetry of the 
wavefunction $\Phi_{i,j}$ when two electrons are interchanged. 
The corresponding energies of $\Psi_{\pm}$ are given by 

\begin{equation}
E_\pm=\frac{\langle\Psi_\pm|h|\Psi_\pm\rangle} {\langle\Psi_\pm|\Psi_\pm\rangle} =\frac{ \langle\Phi_{i,j}|h|\Phi_{i,j}\rangle \pm \langle\Phi_{j,i}|h|\Phi_{i,j}
\rangle} {\langle\Phi_{i,j}|\Phi_{i,j}\rangle \pm \langle\Phi_{j,i}|\Phi_{i,j}
\rangle},
\label{Engpm}
\end{equation}

\noindent where $h$ is the Hamiltonian for the H$^-_2$ system,

\begin{equation}
h = \sum^3_{\alpha=1} \bigl( \frac{p^2_\alpha}{2m} - \frac{e^2}{r_\alpha} - 
\frac{e^2}{r'_\alpha} \bigr) + \sum^3_{\alpha < \beta} \frac{e^2}{r_{\alpha\beta
}}.
\label{h2ham}
\end{equation}

\noindent In Eq.~(\ref{h2ham}), $p^2_\alpha /2m$ is the kinetic energy,
 $r_\alpha$ and $r'_\alpha$ are the distances of the $\alpha$th 
electron to the two protons respectively, 
and $r_{\alpha\beta}=|{\bf r}_\alpha - {\bf r}_\beta|$. It is convenient to
split $h$ into two parts, $h=h_0 + h'$, with $h_0$ the Hamiltonian 
without the interaction between the sites i and j,   
\begin{equation}
h_0=\sum^3_{\alpha=1} \bigl( \frac{p^2_\alpha}{2m}\bigr) - \frac{e^2}{r_1} - 
\frac{e^2}{r_2} - \frac{e^2}{r'_3}. 
\end{equation}

\noindent $h_0$ approaches to $h$ at the limit of large  inter-proton 
distance. $h'$ describes the interaction energy between the H$^-$-ion and the
 H atom. 

Substituting $\Phi_{i,j}$ into Eq.~(\ref{Engpm}) in terms of $\phi_i$ 
and $\phi_j$, we find the two numerator terms to be the following combinations
of integrals of $h_0$ and $h'$,
\begin{eqnarray}
\langle\Phi_{i,j}|h|\Phi_{i,j}\rangle &=& 3(E_0+E_1)-3(E_0 a_2+b_2),\\ 
\langle\Phi_{j,i}|h|\Phi_{i,j}\rangle &=& 3(E_0 a_3+b_3)-3(E_0 a_1+b_1), 
\end{eqnarray}

\noindent where $E's$ and $a's$  are defined as,
\begin{eqnarray}
E_0 &=& \langle (12)_i 3_j|h_0|(12)_i 3_j \rangle ,\quad
E_1 = \langle (12)_i 3_j|h'|(12)_i 3_j \rangle , \nonumber\\
a_1 &=& \langle (12)_i 3_j|1_i (23)_j \rangle ,\qquad ~~
b_1 = \langle (12)_i 3_j|h'|1_i (23)_j \rangle ,\nonumber\\
a_2 &=& \langle (12)_i 3_j|(23)_i 1_j \rangle ,\qquad ~~
b_2 = \langle (12)_i 3_j|h'|(23)_i 1_j \rangle ,\nonumber\\
a_3 &=& \langle (12)_i 3_j|3_i (12)_j \rangle ,\qquad ~~
b_3 = \langle (12)_i 3_j|h'|3_i (12)_j \rangle .
\label{integrals}
\end{eqnarray}

\noindent For simplicity, we define 
$|(12)_i 3_j \rangle \equiv |\psi_i (1,2)\phi_j (3)\rangle$. Note that
$E_0$ is simply the sum of the energies of the independent H$^-$ ion and
H-atom. Eq.~(\ref{Engpm}) now becomes,
\begin{equation}
E_\pm=E_0+\frac{E_1-b_2 \pm (b_3-b_1)}{(1-a_2) \pm (a_3-a_1)}.
\label{Engpm1}
\end{equation}

\noindent These integrals are calculated numerically for a few values of the
inter-proton distance $d$ and are listed in Table~\ref{int}.

\begin{figure}[htbp]
\vspace{0.5cm}
\centerline{\psfig{figure=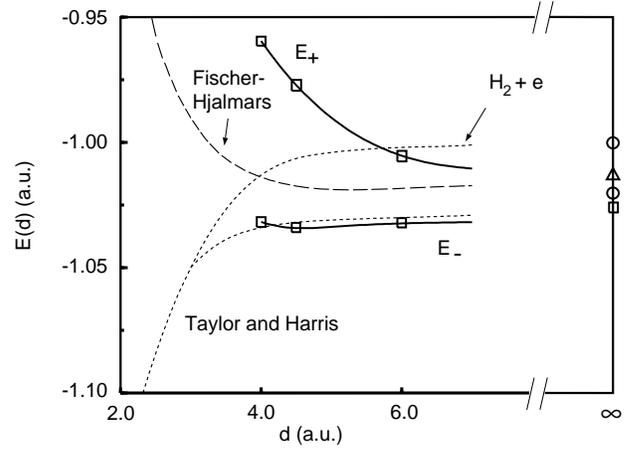,height=6cm}}
\vspace{0.5cm}
\caption{Ground state energies ($E_{-}$, odd parity) and lowest energies of
even parity state ($E_{+}$) in a H$^-_2$ ion as a function of the two proton
distance $d$. The lower (upper) solid line shows the present result for the
bonding energy $E_{-}$ (antibonding energy $E_+$) of the H$^-_2$ ion. 
Dashed line is the $E_{-}$ from Fischer-Hjalmars \protect \cite{Fischer}. The 
lower dotted line is the $E_{-}$ estimated by Taylor and Harris \protect 
\cite{Taylor}. The upper dotted 
line shows the ground state energy of a H$_2$ molecule and 
a free electron for comparison. At $d \rightarrow \infty$, the energy is the 
sum of a H$^-$ ion and a
H atom, shown with a square from the present calculation, a triangle from 
Fisher-Hjalmars and the lower circle from Taylor and Harris. The upper circle
is the energy of two independent H atoms and a free electron.
\label{Eminus} }
\end{figure}

 For H$_2^-$ ion, the ground state is of odd parity. The energy 
of H$_2^-$ ion we 
obtained in the present approach compares favorably with the best
estimates reported previously using different methods \cite{Taylor,Fischer}. 
In Fig.~\ref{Eminus}, we plot our values of $E_{\_}$ as a function of 
inter-proton distance $d$. Fischer-Hjalmars \cite{Fischer} used a similar 
approach, but ignored the correlation term in the Chandrasakhar wavefunction 
probably due to the limited computational power in the 60's. 
Our energies are much lower than theirs. This indicates that the electron 
correlation has a significant contribution
to the ground state energy of the H$^-_2$ ion. It
comes as a surprise that our result is even somehow better than that of Taylor
 and Harris for $d>4$ a.u.. Taylor and Harris \cite{Taylor} used a 
rather complicated wavefunction which involves many linear and nonlinear
variational parameters. Their parameters are distance dependent and variational
procedures were carried out for each  of $d$. The comparison with their 
result indicates that the Chandrasakhar wavefunction that we used is very 
successful on capturing the most essential physics in a single H$^-_2$ ion. 
This also justifies our estimate for the electron hopping integrals in the 
H$^-$ ion lattice. The present calculations give a higher ground state energy 
for $d<4$ a.u. than that of Taylor and Harris. This is understandable.
For smaller distances $d$, our method constructing a wavefunction based on 
individual H$^-$ and H orbitals becomes poor. 
However, the distance between hydrogen ions in 
lanthanum hydrides is always greater than 4 a.u., the region of our interest. 
Therefore we are confident in our estimate of the hopping integral. 

Using Eqs.~(\ref{td}) and (\ref{Engpm1}), we find that the hydrogen
hopping integral $t$ can be 
written in terms of the integrals in Eq.~(\ref{integrals}) as,

\begin{equation}
t=- \frac{(E_1-b_2)(a_1-a_3)-(b_1-b_3)(1-a_2)}{(1-a_2)^2-(a_1-a_3)^2},
\end{equation}

\noindent giving values for the hopping integral between the neighboring 
H$_{\rm tet}$ and H$_{\rm oct}$ (see Fig.~\ref{LaH3}), 
$t_2 = -0.748$ eV ($d=4.58$ a.u.), and between 
the two nearest H$_{\rm tet}$ atoms, $t_1=-0.523$ eV ($d=5.29$ a.u.). 

In the above calculations, we focus on the H$^-_2$ ion, and have neglected the 
Madelung potential of other H$^-$ ions and metal ions in the lattice. These
ions create an
electric field on the H$^-_2$ ion under consideration, which reduces the
amplitudes of hopping integrals. This effect is often called a crystal field 
effect, and will be discussed next.


\subsection{Crystal field effect}
In the above estimation for the hopping integrals,  we have neglected the 
La$^{3+}$ ions as well as surrounding H$^-$ ions. In an ionic picture, 
the La$^{3+}$ and H$^-$ ions generate a crystal field at each H site. Therefore
the Hamiltonian in Eq.~(\ref{h2ham}) should be modified to include the Coulomb
interactions between the electrons in the H$^-_2$ ion under
consideration and all other H$^-$ ions and La$^{3+}$.
We will treat the crystal field effect as a perturbation, whose
Hamiltonian is given by 

\begin{equation}
 h''=\sum_{{\bf R}} -\frac{Z({\bf R})e^2}{|{\bf r}_2-{\bf R}|},
\label{hham}
\end{equation}

\noindent where the sum runs over the La$^{3+}$, and all other H$^-$ ions. 
$Z({\bf R})e$ represents the charge of an ion located at ${\bf R}$ ( 
$Z({\bf R})=3$ for La$^{3+}$ ion, and $Z({\bf R})=-1$ for H$^-$ ion). Then 
the first order correction to the energies $E_\pm$ is given by

\begin{equation}
\delta E_\pm=\frac{E'_1 \mp b'_1}{1 \mp a_1},
\label{deng}
\end{equation}

\noindent with 

\begin{equation}
E'_1= \langle (12)_i 3_j|h''|(12)_i 3_j \rangle,~~ 
b'_1=\langle (12)_i 3_j|h''|1_i (23)_j \rangle.
\end{equation}

\noindent Accordingly, the change of the hopping 
integral, $\delta t=\left(\delta E_{-}-\delta E_{+}\right)/2$ is found to be 

\begin{equation}
\delta t= -\frac{E'_1 a_1-b'_1}{1-a^2_1}.
\end{equation}

\begin{figure}[htbp]
\vspace{0.5cm}
\centerline{\psfig{figure=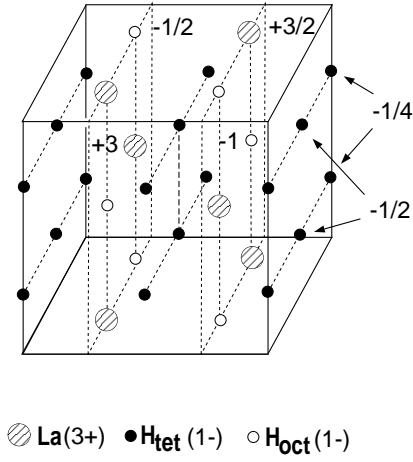,height=6cm}}
\vspace{0.5cm}
\caption{A cubic cell with the center of H$_2^-$ as the origin is shown. 
Total charges of ions enclosed in the cell are neutral. Charges in unit of $e$
on an edge, surface and corner are counted as a quarter, half and eighth of
the original charges respectively. Only a 
few examples are given in the figure. Dotted lines are guides to the eyes. 
\label{cryfld} }
\end{figure}

\noindent In the calculation of $\delta t$, we shall use a summation technique
similar to the usual calculation of the Madelung constant to keep the charge 
neutrality of the summed ions. For a reason to be explained later, we consider
$\delta t_1$, the change of the hopping integral between the nearest 
neighbor (n.n.) tetrahedral 
hydrogens.  We consider the midpoint of the 
H$_2^-$ in question as the origin (Fig.~(\ref{cryfld})), 
and calculate $\delta t_1$ from the 
contributions of all the other ions within a 
cubic cell centered at the origin. As the size of the cubic cell increases, 
$\delta t_1$ changes monotonically and saturate. In Table \ref{crysfield},
we list $\delta t_1$ as the size of the cubic cell
increases. $\delta t_1$ starts to converge for a $4\times 4\times 4 a^3_0$ 
cubic cell and reaches a value $\delta t_1=0.230$ eV at $6\times
6\times 6 a^3_0$ (See Table \ref{crysfield}).
 
Note that $t_1$ is negative , and $\delta t_1$ is positive. The crystal 
field reduces the magnitude of the hopping integral. This is consistent with 
our intuition. The 
net effect of surrounding ions is from positive charged ions, because of
the charge neutrality (H$^-_2$ has a net charge of $-e$). The positive charged
 ions (La$^{3+}$) attract the 
outer electron of a H$^-$ ion, and reduce its hopping amplitude to
a neighboring hydrogen.

We thus estimate the hopping integral in the presence of the crystal field, 
to be $\tilde{t}_1 = -0.293$ eV. This is about 60\% of the value without the 
crystal field. We see that the 
crystal field significantly reduces the hopping integral. However, the 
estimate for $t_2$ is more complicated because of the asymmetry of the crystal
with respect to the tetrahedral and octahedral hydrogen atoms.  Assuming the 
same percentage reduction for $t_2$, we estimate $\tilde{t}_2 = -0.419$ eV. 
Note that our estimate is based on an ideal ionic approach within
which screening 
effects are neglected, so that the actual reduction of the $t$'s is 
expected to smaller.


\section{Microscopic Model for metal hydrides}\label{micmodel}


\subsection{Model Hamiltonian}
Having a  better understanding of the electron correlation in H$^-$ and its 
effect on the electron hopping matrix between the H$^-$ ions, we are 
ready to proceed to an appropriate microscopic model for the
rare earth hydrides. We introduce a large $U$-limit Anderson lattice
Hamiltonian to model the system. In the tight binding representation, the
Hamiltonian, in the second quantization language, is given by:

\begin{equation}
H=H_{h}+H_{La}+H_{mix},
\label{Ham}
\end{equation}

\noindent where the three terms represent the hydrogen, the lanthanum, and 
their hybridization respectively. The hydrogen part is 

\begin{equation}
H_{h}=\sum_{i,s}\epsilon_i^h h^{\dag}_{i,s}h_{i,s} + 
\sum_{\langle i,j\rangle} t_{i,j}^h\left( h^{\dag}_{i,s}h_{j,s} 
+{\rm{h.c.}}\right),
\label{Hh}
\end{equation}

\noindent where $i$ sums over all the occupied H atoms, and 
$\langle i,j\rangle$ neighboring pairs. $h_{i,s}$ is the destruction
operator for an electron of spin $s$ 
on the H site $i$. There is a constraint for 
electrons on each H site $i$,

\begin{equation}
\sum_{s}h^{\dag}_{i,s} h_{i,s} \ge 1. 
\label{constraint}
\end{equation}

\noindent Eq.(\ref{constraint}) is to exclude the empty electron state 
at any hydrogen site. We will discuss this point further below. In 
Eq.(\ref{Hh}), $\epsilon_i^h$ is the atomic energy of the outer electron at
site $i$, which is $-0.7$ eV in free space and will be modified in crystal. 
$\epsilon_i^h=\epsilon_t$ at the tetrahedral site, and 
$\epsilon_i^h=\epsilon_o$ at the 
octahedral site. $t_{ij}^h$ are the hopping integrals between two H sites as
 estimated in section~\ref{Hion}. The lanthanum part of the Hamiltonian is 
given by

\begin{equation}
H_{La}=\sum_{i,\alpha,s}\epsilon_{\alpha}^{La} d^{\dag}_{i,\alpha,s}d_{i,\alpha,s} + 
\sum_{\stackrel{\scriptstyle \langle i,j\rangle}{\alpha,\beta,s}} t_{i,j}^{\alpha,\beta}\left( d^{\dag}_
{i,\alpha,s} d_{j,\beta,s} +{\rm{h.c.}}\right),
\label{HLa}
\end{equation}

\noindent where $d_{i,\alpha,s}$ destroys an electron of orbital $\alpha$ and 
spin $s$ on the La-site $i$. We shall only include the five La-$5d$
orbitals labeled by $\alpha$. They are closest to the chemical
potential and therefore most relevant. 

The La-$6s$ orbital will be
neglected since its energy levels are high above the fermi
energy. $\epsilon_\alpha^{La}$ denotes the atomic energy of orbital
$\alpha$ at the the La-site. $t_{i,j}^{\alpha,\beta}$ is the hopping
integral between orbital $\alpha$ at site $i$ and orbital $\beta$ at
site $j$. 

$H_{mix}$ describes the electron hybridization between H and La sites,
\begin{equation}
H_{mix}=\sum_{i,j,\alpha,s} V_{i,\alpha,j}
\left(d^{\dag}_{i,\alpha,s} h_{j,s} +{\rm{h.c.}}\right), 
\label{Hmix}
\end{equation}

\noindent where $V_{i,\alpha,j}$ is the hopping integral between La site
$i$ of orbital $\alpha$ and H-site $j$.

We now discuss on the physical meaning of the constraint 
Eq.~(\ref{constraint}).
As we examined in the previous section, the two electrons in the H$^-$ ion are
very different in energy. The outer electron has a binding energy 0.7 eV,
while the binding energy for the inner one is 13.6 eV in free space. At low
energies, the inner electron is always occupied, and only the outer
electron is mobile. The constraint (\ref{constraint}) is a mathematical 
description of this physics. If we define a H$^-$ ion as a vacuum, namely a 
fully filled 1$s$ shell, then a neutral H atom is a single-hole state, 
and H$^+$ is
a double-hole state. The constraint Eq.~(\ref{constraint}) prohibits a 
doubly occupied hole configuration at any H site. Our model
(\ref{Ham}) is a large
$U$-limit Anderson lattice model, where $U$ is the Coulomb repulsion between
two holes on the same H site.

The important difference between inner and outer electrons in hydrides was also
discussed by Eder $et$ $al$. \cite{Eder}. These
authors describe hydrogen as a breathing atom, whose radius is much larger for
H$^-$ than for H. Eder $et$ $al$. proposed an Anderson lattice model, where the
hopping integrals depend on electron occupation number on the H site. They
carried out an impurity-like calculation to examine the stability of
the Kondo-like singlet configuration similar to that proposed in our 
earlier work \cite{Ng}.

We now apply Eq.~(\ref{Ham}) to LaH$_x$. In LaH$_x$ the lanthanum sites are 
always fully occupied while the occupation of hydrogen sites depends on $x$. 
The properties of Hamiltonian 
(\ref{Ham})-(\ref{Hmix}) also depends on the parameter values. The
effective hopping integrals $t_{i,j}^h$ between two H sites have been
estimated in section~\ref{Hion}. The other parameters may be extracted 
from the local density approximation calculations, which will be described in 
the next section.

The Hamiltonian (\ref{Ham}) with the local constraint is a many-body 
problem. We
will solve the model for different values of the H concentration $x$ 
using the Gutzwiller method \cite{Gutzwiller,Vollhardt},
which has been well established in study of heavy electron and mixed valence
compounds. 


\subsection{Parameters fitting from LDA calculations}\label{secLDA}

The LDA does not treat the electron correlation
properly. However we expect the LDA to give reasonably good estimates for other
parameters in the hydrides. We carried out LDA calculations for
LaH$_3$, used a tight 
binding model to fit to the LDA results
and extracted parameters for model (\ref{Ham}). For simplicity,
we consider only hopping integrals between the nearest neighbor (n.n.)
tetrahedral H-sites 
($t_2$), between the neighboring H$_{\rm tet}$ and H$_{\rm oct}$ sites ($t_1$),
between the n.n. H and La sites, and between the n.n. and the next n.n. La 
sites. Since the lattice constant changes insignificantly for $3\leq x\leq 2$
in LaH$_x$, we shall assume the parameters remain unchanged.

\begin{figure}[htbp]
\vspace{0.5cm}
\centerline{\psfig{figure=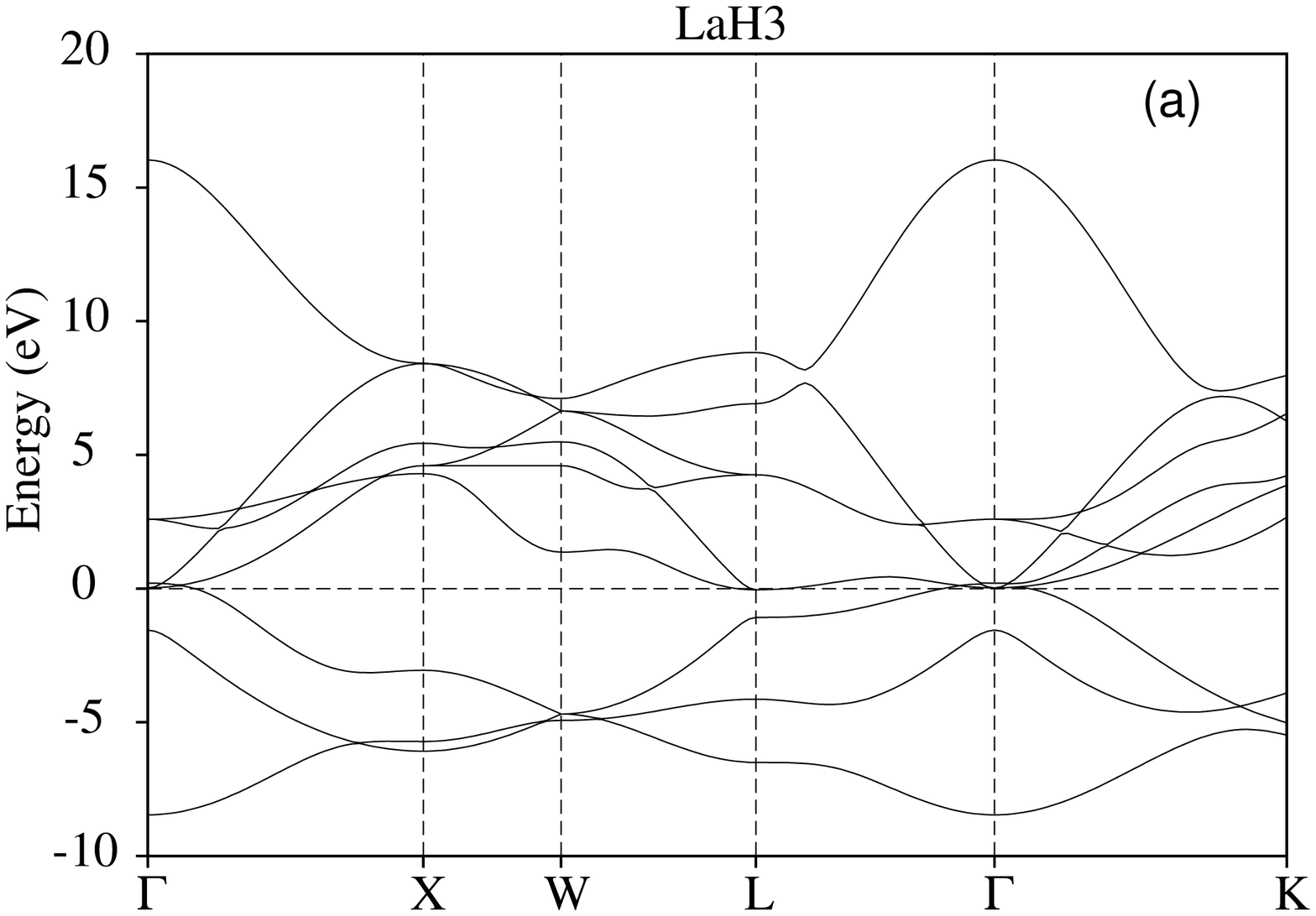,height=5.5cm}}
\centerline{\psfig{figure=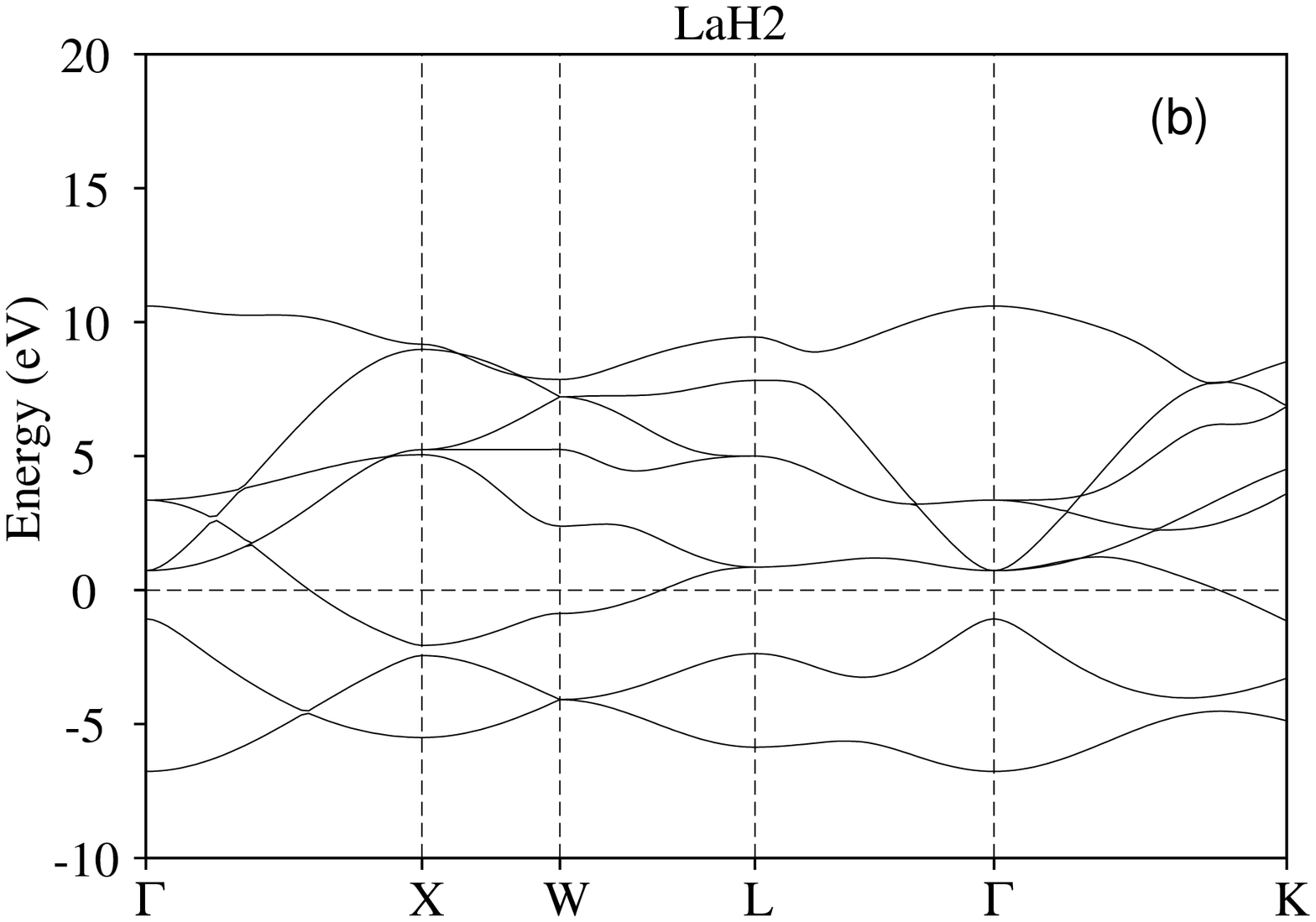,height=5.5cm}}
\vspace{0.5cm}
\caption{Band structures of (a) LaH$_2$ and (b) LaH$_3$ from local density
approximation calculation. The Fermi energy in LaH$_3$ is 
between the top
of the H bands (the lower 3 bands) and the bottom of the La bands (upper 5
bands). $\Gamma=(0,0,0)$, X$=(2\pi,0,0)$, W$=(2\pi,\pi,0)$, L$=(\pi,\pi,\pi)$,
K$=(3\pi/2,3\pi/2,0)$, in unit of 1/$a_o$, are the high symmetry points in 
crystal momentum space.
 \label{vladband}}
\end{figure}

\begin{figure}[htbp]
\vspace{0.5cm}
\centerline{\psfig{figure=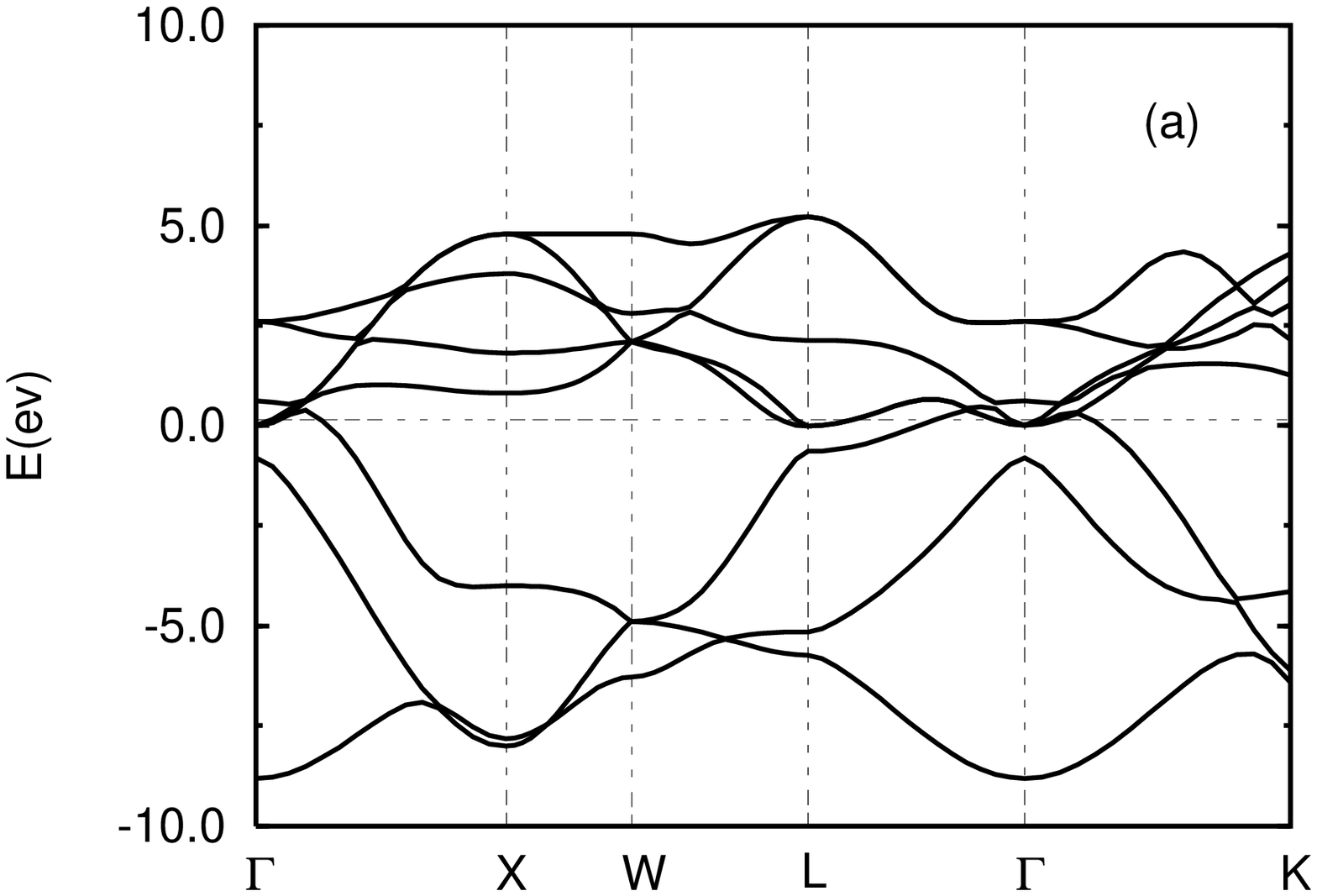,height=5.5cm}}
\centerline{\psfig{figure=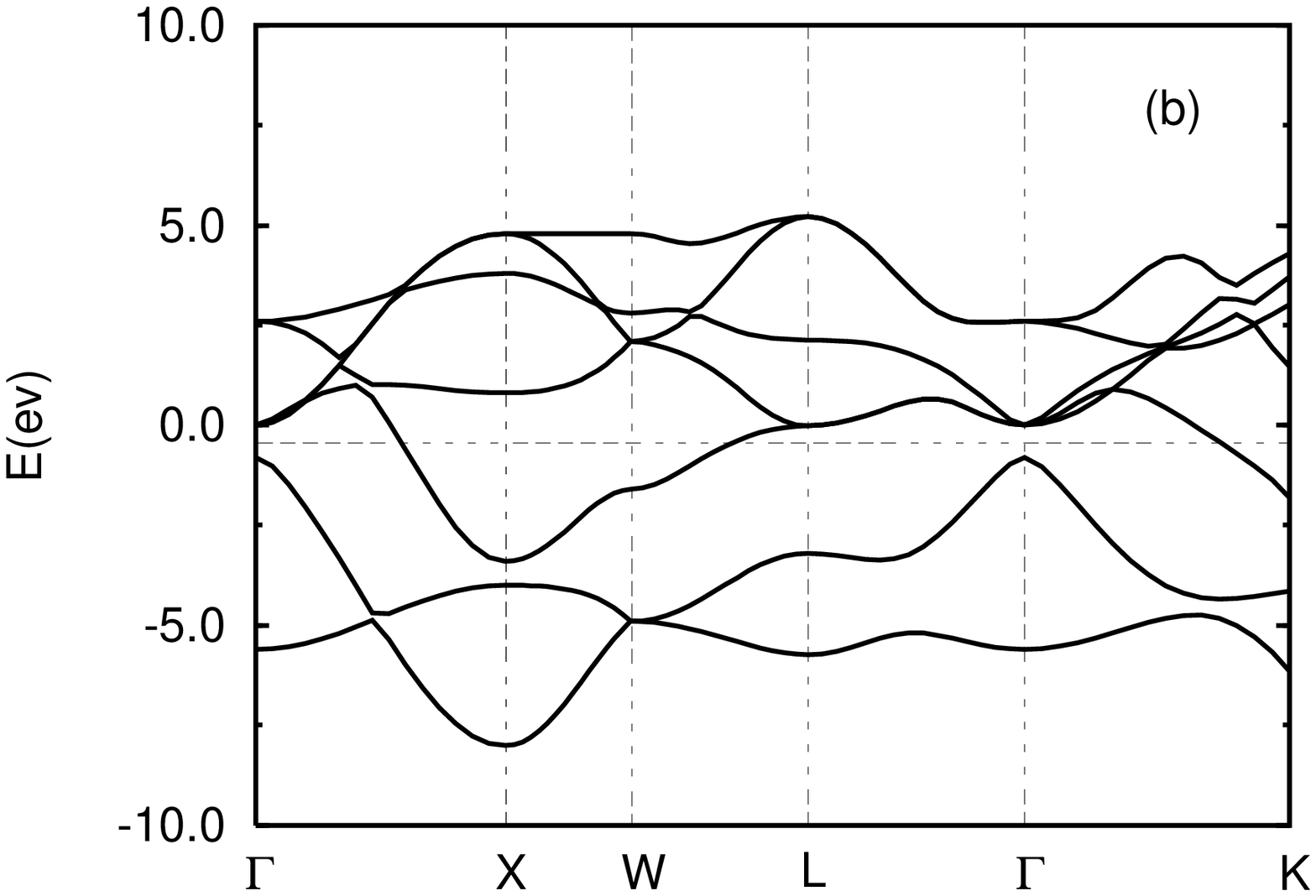,height=5.5cm}}
\vspace{0.5cm}
\caption{The fitted band structures from the tight binding
model with parameters listed in Table \protect\ref{parameter} for (a) LaH$_3$
and (b) LaH$_2$.
\label{fitband}}
\end{figure}

In Fig.~\ref{vladband} we show our LDA results for the band structure of 
LaH$_3$ and LaH$_2$. Other LDA calculations gave similar results 
\cite{Dekker,Wang}. We use a tight binding model, identical to Eq.~(\ref{Ham})
but with no constraint Eq.~(\ref{constraint}), to fit Fig.~\ref{vladband}(a) 
for LaH$_3$. The fitted band structure is plotted in Fig.~\ref{fitband}(a), 
and the fitting parameters are listed in Table \ref{parameter}. In the 
fitting, we pay most attention to those states near the chemical potential. The
same set of parameters fit the LDA bands of LaH$_2$ as well, see 
Fig.~\ref{vladband}(b) and Fig.~\ref{fitband}(b).

Since all $s$-$s$, $s$-$d$ and $d$-$d$ orbital bondings can be expressed as a 
linear combinations of $\sigma$-bonds (angular momentum along the bonding 
axis $m=0$), $\pi$-bonds ($m=1$) and $\delta$-bonds ($m=2$), the hopping 
integrals $t_{i,j}^{h} $, $t_{i,j}^{\alpha,\beta}$ and $V_{i,\alpha,j}$ can be 
represented by a few
bonding integrals. They are given by

\begin{eqnarray}
 t_{i,j}^{h} &=& V_{ss\sigma}^{h-h},\nonumber \\
t_{i,j}^{\alpha,\beta} &=& a_{\alpha,\beta}(l,m,n)V_{sd\sigma}^{La-h},
\nonumber \\
V_{i,\alpha,j} &=& b_{\alpha}^{(1)}(l,m,n)V_{dd\sigma}+b_{\alpha}^{(2)}(l,m,n)
V_{dd\pi}.
\end{eqnarray}

\noindent where $V_{ss\sigma}^{h-h}$ ($V_{sd\sigma}^{La-h}$) is 
$\sigma$-bonding integrals between H $1s$ and H $1s$ (La $5d$) orbitals, while
$V_{dd\sigma}$ and $V_{dd\pi}$ are $\sigma$- and $\pi$-bonding integrals of 
La $5d$ to $5d$ orbitals, respectively. The contribution from $\delta$-bonding
is generally small and is ignored. The coefficients $a_{\alpha,\beta}$ and
$b_{\alpha}$'s are functions of the direction cosines $(l,m,n)$ from site $i$
to site $j$. These coefficients have been derived by Slater and Koster
\cite{Slater} based on the symmetry. The number of fitting 
parameters is now reduced since there are only a few bonding
integrals. It is worth to note that the H hopping integrals obtained
from LDA, $V_{ss\sigma}^{o-t}=-0.79$ and $V_{ss\sigma}^{t-t}=-0.4$,
are larger than that calculated from the local model,
$\tilde{t_2}=-.419$ and $\tilde{t_1}=-0.293$. It justifies that the
electron correlation plays an important role in the H hopping integrals.
Therefore, we will use $\tilde{t_1}$ and $\tilde{t_2}$
for the H hopping integrals in the future calculations.

Although there are some discrepancies with 
the LDA band structure for high energy La bands, the main concern here 
should be the position of the lowest La bands and the highest H bands 
which are responsible for both the transport and optical properties. 
Bear in mind that our motivation is to reproduce the essential 
qualitative features rather than a very accurate description and 
hence the fitting may not be unique. We have carried out the same 
fitting procedures for Dekker's LDA band calculation \cite{Dekker}. Even 
though it 
generates a different set of parameters, the general features like 
the opening of band gap when 
introducing electron correlation and the localization of the 
vacancy and impurity state, which will be discussed later, 
are consistent with the current set of parameters. Therefore, we are 
confident in the use of this parameter set for our future calculations.


\subsection{Gutzwiller method for the many-body Hamiltonian}\label{secgut}

Following Rice and Ueda \cite{Rice}, we shall use the Gutzwiller method to 
study the many-body Hamiltonian (\ref{Ham}).
The method is a variational approach. The constraint condition of
Eq.~(\ref{constraint}), to exclude state at any H site, is replaced by
a set of renormalization factors.

We consider a variational ground state wavefunction $|\Psi\rangle$ which is of 
the form 
\begin{equation}
|\Psi\rangle=P|\Psi_o\rangle,
\end{equation}

\noindent where $|\Psi_o\rangle$ is a state of the tight binding Hamiltonian 
with no 
constraint, and $P$ is the projection operator to project all empty 
electron states on each H site, 
\begin{equation}
P=\prod_i n_{i,\uparrow}^h n_{i,\downarrow}^h,
\end{equation}

\noindent where $i$ runs over all H sites and 
$n_{i,s}^h=h_{i,s}^{\dag}h_{i,s}$.

The ground state energy is given by 
\begin{equation}
 E_g=\langle H\rangle=\frac{\langle\Psi\left|H\right|\Psi\rangle}
{\langle\Psi|\Psi\rangle}.
\end{equation}

\noindent To calculate $\langle H \rangle$, we use the Gutzwiller method to 
compute the average values of each term in $H$ in the state $|\Psi\rangle$. The
Gutzwiller method is a static approximation, and the average of a local
operator $Q$ in the state $|\Psi\rangle$ is related to its corresponding 
average in the $|\Psi_o\rangle$ by a renormalized numerical factor $g_Q$,
namely, 
\begin{equation}
\langle Q \rangle=g_Q \langle Q \rangle_o,
\end{equation}

\noindent where $\langle Q \rangle$ and $\langle Q \rangle_o$ are the average
in the states $|\Psi\rangle$ and $|\Psi_o\rangle$ respectively.

In the present case, we have 
\begin{eqnarray}
\langle h_{i,s}^{\dag} h_{j,s} \rangle &=&\sqrt{g_i g_j} 
\langle h_{i,s}^{\dag} h_{j,s} \rangle_o,\nonumber\\
\langle d_{i,\alpha,s}^{\dag} h_{j,s} \rangle &=&\sqrt{g_j} 
\langle d_{i,\alpha,s}^{\dag} h_{j,s} \rangle_o.
\end{eqnarray}

The Gutzwiller factor $g$ is calculated by counting the number of possible
configuration in $|\Psi\rangle$ and in $|\Psi_o\rangle$,
respectively. The value of $g_i$ is 

\begin{equation}
g_i = \frac{1-n^{hole}_i}{1-n^{hole}_i/2},
\label{gi}
\end{equation}

\noindent where $n^{hole}_i$ is the occupation number of holes at the H site 
$i$. Here we define a hole as an electron vacancy in a H$^-$ ion and therefore

$$
n^{hole}_i=2-\sum_{i,s} h^{\dagger}_{i,s} h_{i,s}.
$$

The constraint Eq.~(\ref{constraint}) now disallows doubly occupied hole
configurations on H sites. This leads to an effective Hamiltonian
for the lanthanum hydrides,

\begin{equation}
H_{eff}=H'_h+H_{La}+H'_{mix}-\mu \sum_{i,s} h^{\dag}_{i,s}h_{i,s}.
\label{Heff}
\end{equation}

\noindent In the above equation, $H_{La}$ is given by Eq.(\ref{HLa}), and 
$H'_h$ and $H'_{mix}$ have the same form of $H_h$ and $H_{mix}$ in 
Eq.(\ref{Hh}) and (\ref{Hmix}). The constraint condition is released, and the
hopping integrals are renormalized to values

\begin{eqnarray}
t_{i,j}^h &\rightarrow& \sqrt{g_i g_j} t_{i,j}^h, \nonumber\\
V_{i,\alpha,j} &\rightarrow& \sqrt{g_j} V_{i,\alpha,j}.
\end{eqnarray}

\noindent H hopping integrals $t_{i,j}^h$ are the values obtained from
our local model including the crystal field effect. Furthermore, the 
atomic energies $\epsilon_i^h$ are shifted to
$\epsilon_i^h-\epsilon_b$, where $\epsilon_b$ is the binding energy of the
outer electron in a H$^-$ ion.  We use the free space value $\epsilon_b=0.7$eV.
Note that the outer electron in LDA is unbound \cite{Gunnarrson}. $\mu$
Eq.~(\ref{Heff}) is an effective "chemical potential" \cite{Rice}. The value 
of $\mu$ is determined by minimizing the total energy of the system

\begin{equation}
E_g=\langle H_{eff}+\mu \sum_{i,s} h_{i,s}^{\dag} h_{i,s} \rangle_o.
\end{equation}

\noindent In the lanthanum hydrides, the total energy is found to be 
insensitive to values of $\mu\simeq 0$, so we set $\mu=0$. This is 
primarily due to the fact that $\epsilon^h$ is lower than $\epsilon^\alpha$. 

To compare with heavy fermion systems, we may perform an electron-hole
mapping for the lanthanum hydride model Hamiltonian, $\epsilon^h \rightarrow
-\epsilon^h$, $\epsilon^\alpha \rightarrow -\epsilon^\alpha$, and the 
constraint is to project out doubly occupied hole states on H sites. In the 
hole notation, the "f" level in lanthanum hydrides is above the conduction
bands. Therefore the hydrides are in the mixed valence, instead of the Kondo 
limit of the Anderson lattice model.

$H_{eff}$ is a single particle Hamiltonian, and can be readily solved by 
analytic or numerical means. The Gutzwiller factor $g_i$, given by 
Eq.(\ref{gi}), satisfies a self-consistent equation for the electron
occupation at H site $i$,

\begin{equation}
n_i^h=\langle n_i^h \rangle=\langle n_i^h \rangle_o.
\end{equation}

Since $0 \leq g \leq 1$, the hopping integrals are 
then renormalized to smaller values. Hence the hydrogen band width is 
expected to be further reduced as a consequence of the correlation.In general
$g$ is position dependent.
In the periodic cases such as in LaH$_2$ and LaH$_3$, $g$ depends only 
on whether it is on the tetrahedral or octahedral site. For the inhomogeneous
impurity states,  $g$ are strongly site dependent. 

Note that this
occupation related correlation is different from the intrasite
electron-electron 
correlation we discussed in section \ref{Hion}. There the interaction between
electrons on the same H$^-$ ion was considered, and the result is 
the correlated wavefunction of two electrons on the same site. Here in this
section we consider the correlation  effect on the intersite hopping, which is
a constraint on the every H-site.


\section{Electronic structure of LaH$_2$ and LaH$_3$}

We are now in position to calculate the electron spectrum of lanthanum 
hydrides. In this section, we consider LaH$_2$ and LaH$_3$. 

\begin{figure}[htbp]
\vspace{0.5cm}
\centerline{\psfig{figure=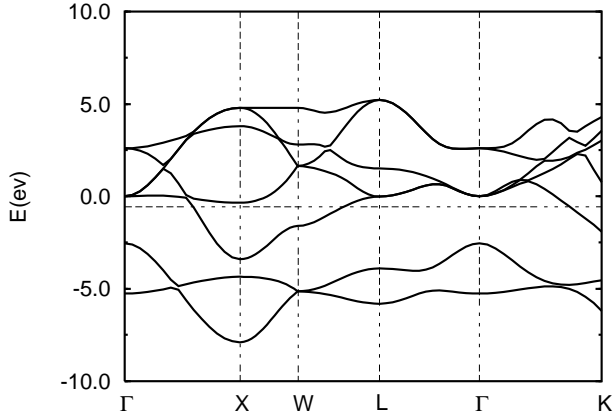,height=5.5cm}}
\vspace{0.5cm}
\caption{The conduction and valence bands of LaH$_2$ calculated from model
Hamiltonian (\protect\ref{Ham}) by using Gutzwiller method. 
\label{gut1}}
\end{figure}

In the case of LaH$_2$, all tetrahedral hydrogen sites are occupied, and all
octahedral H sites  are unoccupied. Because of 
 the equivalence between the two tetrahedral H atoms in the unit cell, we
need to introduce only one Gutzwiller factor, $g_t$. The energy spectrum, or
the band structure, obtained from the $H_{eff}$, is plotted in 
Fig.~\ref{gut1}. The two lower bands are primarily H$^-$ bands, and are 
fully filled, leaving one electron per unit cell in the bands of 
primarily La-$5d$ characters. The Fermi energy is within the La-$5d$ 
conduction bands, so that LaH$_2$ is a metal as  expected. The 
correlations in LaH$_2$ do not change this qualitative property.

\begin{figure}[htbp]
\vspace{0.5cm}
\centerline{\psfig{figure=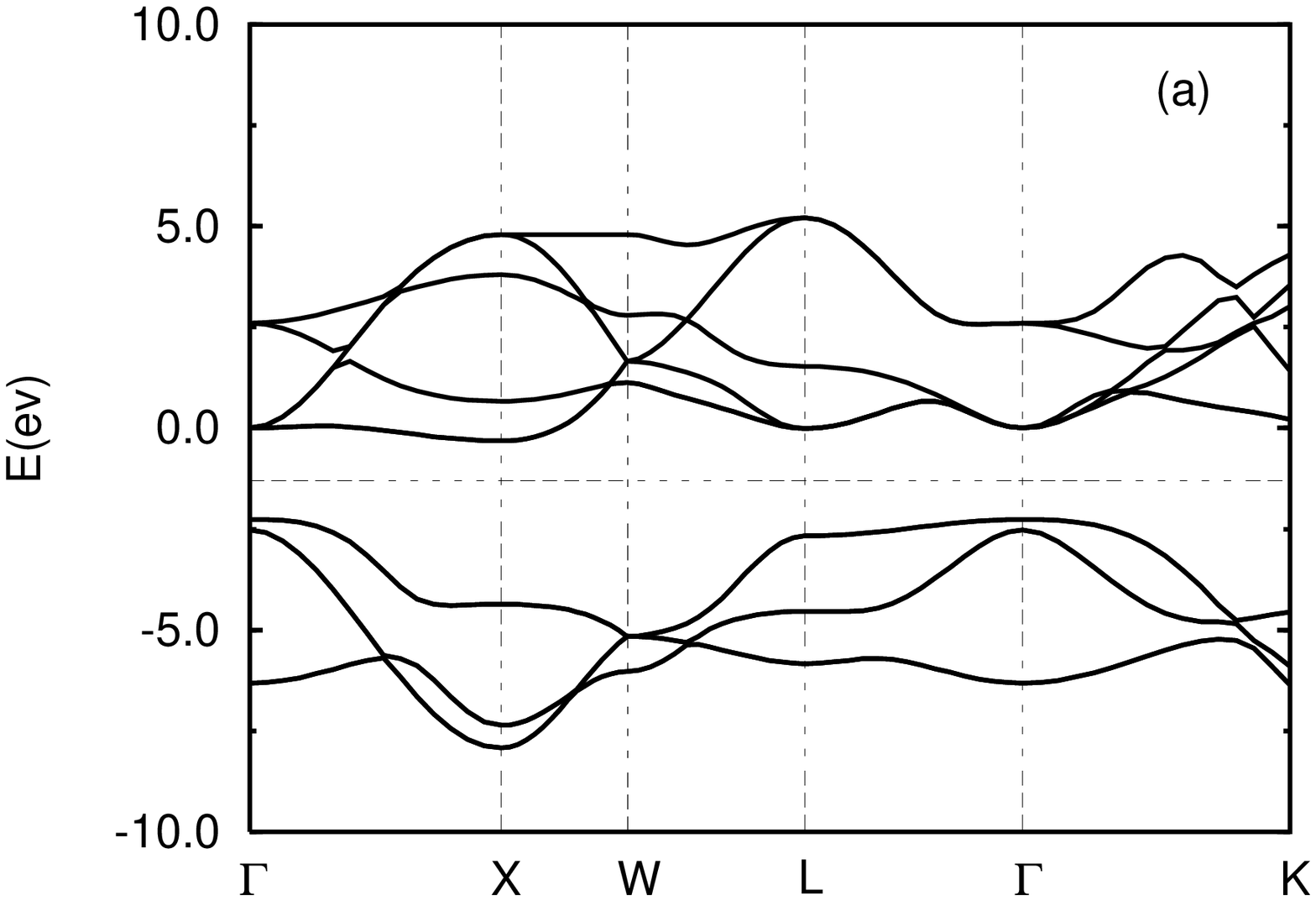,height=5.5cm}}
\vspace{0.5cm}
\centerline{\psfig{figure=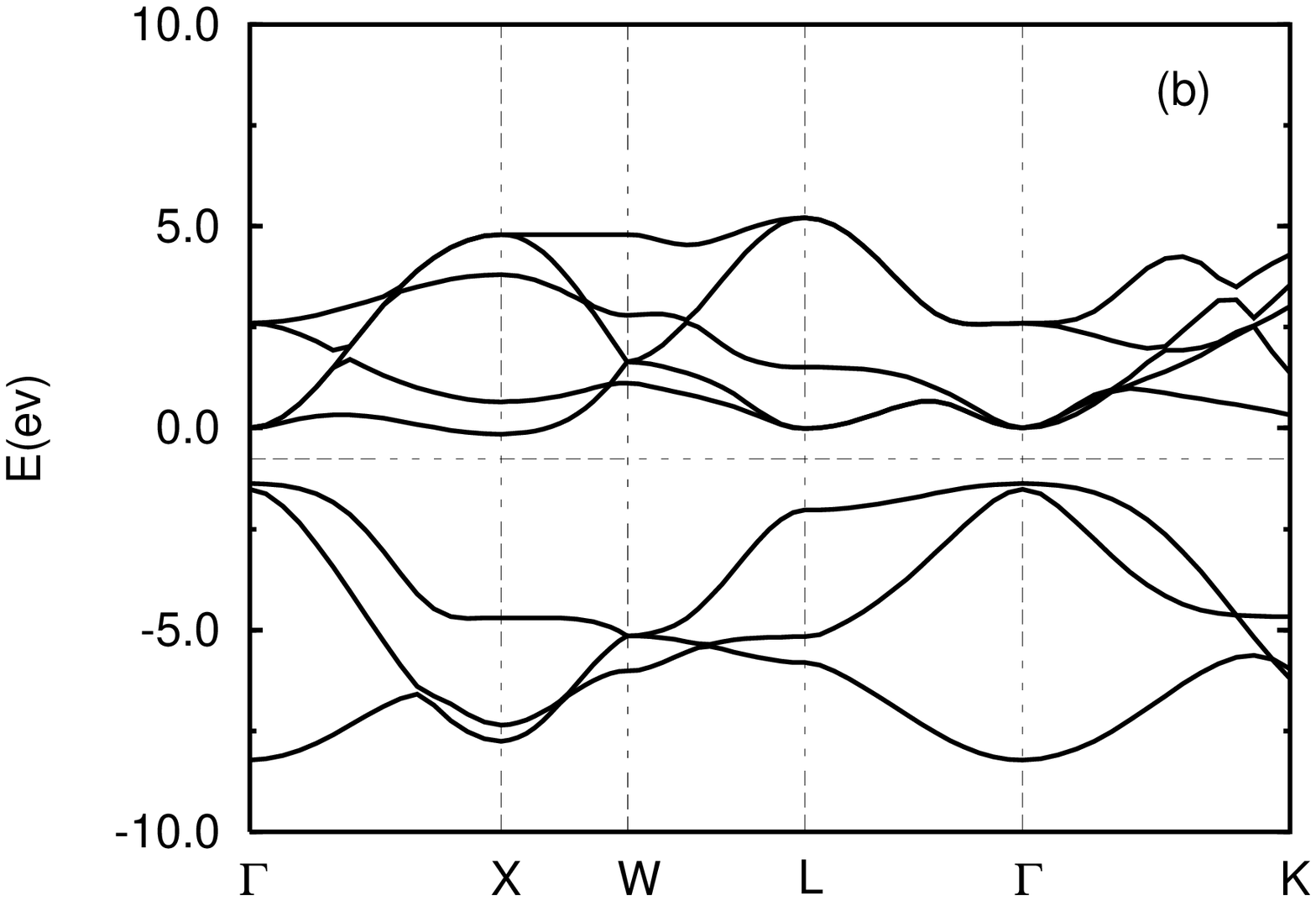,height=5.6cm}}
\vspace{0.5cm}
\caption{Band structure of LaH$_3$ of model Hamiltonian (\protect\ref{Ham}) 
using
Gutzwiller approximation. In (a) the crystal field is included in estimating
the H-H hopping integrals. In (b) the crystal field effect is not included.
The actual result is expected to be between (a) and (b). 
\label{gut2}}
\end{figure}

In LaH$_3$, all octahedral H sites  are also occupied. We use two independent
Gutzwiller factors $g_t$ and $g_o$ corresponding to the tetrahedral and
octahedral sites respectively. The Gutzwiller factors are found to be 
$g_t=0.78$, $g_o=0.70$, which indicates the reduction of the hopping 
parameters between two H sites and
between H and La sites. The electronic structure for LaH$_3$ is plotted
in Fig.~\ref{gut2}.

We find that LaH$_3 $ is an insulator. The optical gap is found to be 1.5~eV
or 2.1~eV depending on whether the crystal field is included in the estimate of
the H-H hopping integrals as discussed in section II.
The opening of the energy gap at the chemical potential is primarily 
due to the strong electron correlations in the H$^-$ ions, which reduces 
the H$^-$ band width. The large electron-electron Coulomb repulsion 
on the H$^-$ ion also restricts the electronic state to exclude the
empty electron state at any H-site, which further reduces the H$^-$
band width. The energy of the conduction band at X-point seems too low, and is
presumably due to the inaccuracy in the tight-binding fit, see 
Fig.~\ref{vladband}(a) and \ref{fitband}(a).

The present method is an improvement of our previous work \cite{Ng}. In that
paper, we focused on the H bands and neglected the La bands. That is
to assume that all the three lower bands, which is of primarily H-$1s$
character and mixed  with the La-$5d$ orbitals, are purely H
bands. Mathematically, the previous model is equivalent to
the effective Hamiltonian Eq.~(\ref{Heff}) but without the
renormalization factors (i.e. $g_i=1$). The more sophisticated method
presented here predicts a larger gap, closer to the observed value.

\begin{figure}[htbp]
\vspace{0.5cm}
\centerline{\psfig{figure=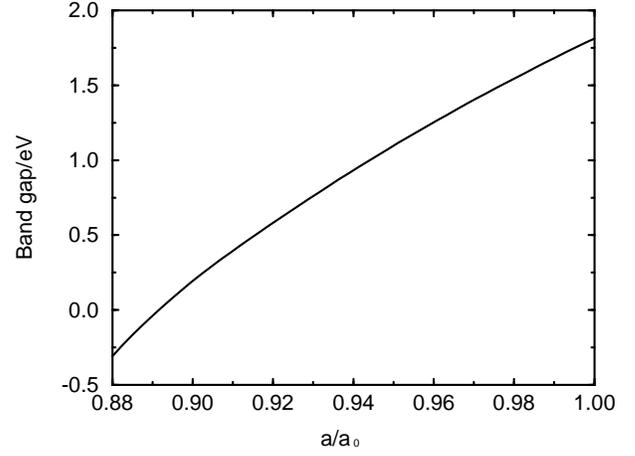,height=6cm}}
\vspace{0.5cm}
\caption{Band gap of LaH$_3$ is observed to be a monotonic increasing function
 of lattice constant $a$ (in unit of $a_0$). Therefore by exerting large 
enough pressure on the crystal, the band gap will be closed up and the 
originally insulating hydride will become conducting. This transition happens 
when $a \sim 0.89 a_0$. 
\label{gap}}
\end{figure}

It would be interesting to study the band gap in LaH$_3$ 
as a function of the lattice constant $a_o$. As the lattice constant 
decreases, the electron hopping matrix elements between H-ions increase, and 
thus the H$^-$ band width increases, possibly leading to 
the reduction of the gap and eventually to the closing of the gap.
Therefore we expect a transition from insulator to a metal
in LaH$_3$ under pressure.  With the knowledge of the distance
dependence of the hydrogen couplings $t_1$ and $t_2$, crystal field
effect ($\propto 1/a_0$)  
and the orbital coupling parameters ($V_{dd\pi}$ and 
$V_{dd\sigma} \propto 1/a_0^5$, $V_{sd\sigma} \propto 1/a_0^{7/2}$) 
\cite{Harrison}, we can calculate the evolution of the band gap as a function 
of $a_0$. We find that the band gap closes and LaH$_3$ becomes 
metallic when $a_0$ decreases by 11\% to $0.89 a_o$. (Fig.~\ref{gap}). 


\section{Impurity and vacancy states}

In the previous two sections, we argued that electron correlation is the 
missing element from the LDA calculations, which failed to predict
the insulating properties of the trihydrides. By introducing an electron 
correlation on H$_2^-$ ions, we have found that LaH$_3$ 
is an insulator, consistent with experiment. However, more interesting
physics lies in the mechanism of the metal-insulator transition in LaH$_{x}$. 
We can understand the transition by approaching it from either metallic $(x=2)$
and insulating $(x=3)$ end points. We investigate the former 
first.


\subsection{An extra H atom in LaH$_2$}

\begin{figure}[htbp]
\vspace{0.5cm}
\centerline{\psfig{figure=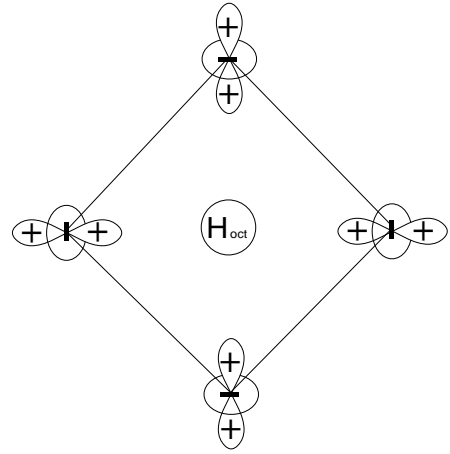,height=6cm}}
\vspace{0.5cm}
\caption{2-D illustration of an impurity state. Six neighboring 
La-$5d$-$e_g$ orbitals (only 4 of them are shown here) are pointing toward 
the extra H$_{\rm oct}$. 
\label{imp2d}}
\end{figure}

The introduction of
a neutral H atom into a H$_{\rm oct}$ site in LaH$_2$ creates a $s$=1/2 
magnetic impurity, which couples to the conduction electron spins.
Let us study this impurity problem more carefully. First we consider the
atomic limit and neglect the hybridization between the H-1$s$ and La-$5d$
orbitals. Then a conduction electron (La-$5d$) moves to the neutral H atom
at H$_{\rm oct}$ site to form a H$^-$ ion. Including the hybridization 
between H and La orbitals, the outer electron in the H$^-$ ion will fluctuate
between the H$_{\rm oct}$ site and the surrounding La sites. Since H$^-$ ion
is a spin singlet of two electrons, the coupling between H$_{\rm oct}$ and La
orbital is antiferromagnetic. The effective Hamiltonian for the extra H 
atoms at H$_{\rm oct}$ is an antiferromagnetic Kondo model,

\begin{equation}
{\cal H}_{\rm imp}\; =\; \sum_{{\bf k},\alpha, s} {\varepsilon_{\alpha}
 ({\bf k}) d^{\dagger}_{{\bf k},\alpha,s}d_{{\bf k},\alpha,s}} + 
J \sum_{i} {{{\bf S}_i} \cdot {{\bf s}_i}}, 
\end{equation}

\noindent where $i$ runs over all the occupied  H$_{\rm oct}$, ${\bf S}$ and
${\bf s}$ are the electron spins of the neutral H atom and of the
conduction electron (annihilation operator $d_{{\bf k},\alpha,s}$ with energy
$\varepsilon_{\alpha}({\bf k})$) states. Because of the symmetry, only the
two $e_g$ La-$5d$ orbitals are coupled to the 1$s$-H$_{\rm oct}$ electron
directly.  The exchange coupling will be large, $J \sim$ 0.7 eV is in
free space~\cite{Kondo}. To estimate $J$ in the hydrides, we use a
local model including six neighboring La atoms around the  H$_{\rm
oct}$ site (Fig.~\ref{imp2d}). The dynamics is given by a $2\times2$
matrix Hamiltonian 

\begin{equation}
\left( \begin{array}{cc}
\epsilon_{oct} & \sqrt{6}V  \\
\sqrt{6}V      & \bar{\epsilon}_d
\end{array} \right),
\label{dynamic}
\end{equation}

\noindent where $\epsilon_{oct}$ is the atomic energy of the 1-$s$ 
H$_{\rm oct}$ and $V=V^{La-o}_{sd\sigma}$ is the hybridization 
between the H$_{\rm oct}$ and each La. $\bar{\epsilon}_d$ is the energy of the 
 combination of the six $5d$-La-$e_g$ orbitals around the H$_{\rm
oct}$, denoted by 

\begin{equation}
|imp\rangle = \frac{1}{\sqrt{6}} \sum_{i} |e^i_g\rangle,
\end{equation}

\noindent where $|e^i_g\rangle$ is the $e_g$ orbitals as shown in 
Fig.~\ref{imp2d} at the La site $i$. Using the Hamiltonian $H_{La}$ in 
Eq.~(\ref{HLa}), we find the energy of $|imp\rangle$ is to be given by

\begin{eqnarray}
\bar{\epsilon_d}&=&\langle imp\left|H_{La} \right|imp\rangle,\\ \nonumber
&=& \epsilon_d'+\epsilon_{Kin}.
\end{eqnarray}

\noindent $\epsilon_{Kin}$ is the kinetic energy of the electron in the 
$|imp\rangle$,

\begin{eqnarray}
\epsilon_{Kin}&=&4 t^{e_g-e_g}\nonumber\\
&=& \frac{V_{dd\sigma}}{4}-3 V_{dd\pi},
\end{eqnarray}

\noindent where $t^{e_g-e_g}$ is the hopping integral between two La-$e_g$
states whose orbitals point to the central H$_{\rm oct}$. From 
Eq.~(\ref{dynamic}), we find the spin-singlet state energy  to be 

\begin{equation}
 \epsilon_{s=0}=\frac{1}{2}\left[\left(\epsilon_{oct}+\bar{\epsilon}_d\right)-
\sqrt{\left(\epsilon_{oct}-\bar{\epsilon}_d\right)^2+24V^2}\right]. 
\label{bondeng}
\end{equation}

\noindent Using the parameters fitted to the LDA, $\epsilon_{oct}=-3.3$~eV,
$\bar{\epsilon_d}=-0.7$~eV, $V=-1.2$~eV, we have $\epsilon_{s=0}=-5.21$~eV. 
This energy should be compared with the chemical potential ($\mu=-0.62$ eV, 
see Fig.~\ref{gut1}) 
for a spin triplet state, in which an electron with the same spin is placed at
the Fermi level conduction band. Therefore, we arrive at a value

\begin{equation}
J=\mu-\epsilon_{s=0}=4.59~{\rm eV}.
\label{JJ}
\end{equation}

In this Kondo description
conduction electrons are captured by the neutral H atoms at 
H$_{\rm oct}$ sites to form tightly bound singlets. In this
description the material with all octahedral sites occupied, LaH$_3$, is 
viewed as a Kondo insulator with a large band gap. Our conclusion here is 
consistent with Eder $et$ $al.$ \cite{Eder}. The actual gap in the
insulator is expected to be reduced from the value in Eq.~(\ref{JJ})
obtained within an impurity calculation. 


\subsection{Localized vacancy state in LaH$_3$}\label{secvac}

Next we examine the transition starting from insulating LaH$_3$. The
removal of neutral H atoms introduces vacancies at octahedral sites
H$_{oct}^V$ which donate an electron to the conduction band.
In a conventional semiconductor such as in the phosphorus-doped silicon
(Si:P), the impurity state is described by an effective mass theory, 
and  the result is a hydrogen-like bound state with a large 
effective Bohr radius $a_B^*$ of order $\sim 100$~{\AA} due to the light 
effective electron mass and the large dielectric constant. 
The critical impurity concentration $\delta_c$ at 
which the system becomes metallic is given by Mott criterion 
$\delta_c \sim (l/a_B^*)^3$, where $l$ is the inter-atomic distance. 
Because of $a_B^* \gg l$, $\delta_c$ is very small ($10^{-3}$ for Si:P). 
The vacancy state in LaH$_{3-\delta}$ is very different, 
however. Experimentally, it is found that the
semiconducting states extend to $\delta_c=0.25$ for LaH$_{3-\delta}$, and
to even larger value for yttrium hydrides~\cite{Huiberts1,Shinar}. 
This suggests a very localized impurity state. 
In our previous short paper we proposed the vacancy state in LaH$_{3-\delta}$ 
to be a very
localized La-$5d$-$e_g$ orbitals centered at the vacancy with $s$-wave 
symmetry, and we used perturbation theory to estimate its energy in support
of our proposal. Here we calculate the lowest vacancy state and further support
our previous vacancy model.

Let us begin with a qualitative argument and a simple estimate. The 
La-$5d$-$e_g$ and H$_{\rm{oct}}$ electrons hybridize 
strongly to form bonding (mainly H) and antibonding (mainly $e_g$) states. 
The underlying physics may be simply described by the local model of a single 
H$_{\rm oct}$ atom surrounded by six La atoms (see last section). The bonding 
state energy is given by Eq.~(\ref{bondeng}), and the anti-bonding state 
energy is 
\begin{equation}
 \epsilon_{ab}=\frac{1}{2}\left[\left(\epsilon_{oct}+\bar{\epsilon}_d\right)+
\sqrt{\left(\epsilon_{oct}-\bar{\epsilon}_d\right)^2+24V^2}\right]. 
\end{equation}

\noindent A vacancy of H$_{\rm{oct}}$ breaks the bonds locally and the $e_g$ 
electron becomes locally non-bonding, which has much lower 
energy than the antibonding $e_g$ states away from the vacancy.  
Therefore, in addition to serving as a positive charge center as in 
conventional n-type semiconductors,  H$_{oct}^V$ creates a potential 
well for the $e_g$ state electron. The latter is non-perturbative, 
and is responsible for the unusual concentration dependence of 
the semiconductor.

Consider a $s$-symmetric octahedral  $e_g$ state ($S$-state hereafter) 
around a  H$_{\rm oct}^V$  as shown in Fig.~\ref{vacancy}. The 
H$_{\rm oct}^V$ vacancy
 breaks the bonds on the octahedron and reduces the antibonding energy
 of the $S$-state. This $S$-state is very localized because
 the neighboring octahedral $s$-state is antibonding and has a much
 higher energy. We can estimate the depth of the
 potential well, $V_0$. The vacancy state (the non-bonding state) energy is
\begin{equation}
E_{vac}=E_{nb}=\bar{\epsilon}_d.
\label{nonbond}
\end{equation}

\noindent Therefore $V_o=E_{ab}-E_{nb}$ is given by
\begin{equation}
V_o=\frac{1}{2}\left[\epsilon_{oct}-\bar{\epsilon}_d+\sqrt{\left(
\epsilon_{oct}-\bar{\epsilon}_d\right)^2+24 V^2}\right].
\end{equation}

\noindent The numerical value of the potential well is $V_o=1.9$~eV.

From Eq.~(\ref{nonbond}), we estimate $E_{nb}=-0.7$~eV, below the lowest 
La-$5d$-$t_{2g}$ band which has energy zero as we can see from 
Fig.~\ref{gut2}(a).

In the above estimate for the energy of the localized vacancy state, we have
not included a Coulomb attraction between the H$_{\rm oct}$ vacancy
and the extra 
electron. This attraction will further lower the vacancy state energy. We
conclude that the H$_{\rm oct}$-vacancy electron 
state is highly localized and it is well within the band gap.
The localized vacancy state here is similar to 
the carbon vacancy state in TiC, where the C-vacancy leads to 
a well localized impurity states \cite{Redinger}. We emphasize that the  
potential well generated by the 
H$_{\rm oct}$ vacancy is primarily short-ranged, different from the ideal 
long-range Coulomb force. The effective  radius of the impurity state  is  
only half of  the lattice constant. 

The existence of localized vacancy state can be verified by a finite size 
calculation for a single H$_{\rm oct}$ vacancy in LaH$_3$, an improvement
of the local model discussed above. The system is 
described by Hamiltonian (\ref{Ham})-(\ref{gi}) with all but one H$_{\rm oct}$
occupied. We use the Gutzwiller method given in section~\ref{secgut} to study 
the vacancy state. The vacancy breaks the translational invariance;
the Gutzwiller factor $g_i$ now depends on the site, and has only
crystal group symmetries with respect to the vacancy state. Because of
its localized nature, we can study the vacancy state in a finite-size
system. The problem then can be solved explicitly using the Gutzwiller
approximation. We consider a small cubic of $3\times3\times3a^3_o$
lattice in the calculations. We choose the vacancy site to be at the center 
of the lattice. An $S$-state around the vacancy site is found, as we expected. 
All the surrounding La atoms have their $5d$ $e_g$ orbitals 
pointing towards the vacancy and have the same weighting factor. The
$t_{2g}$ orbitals, however, have no contribution in the $S$ state.  It
is found that the $S$ state has an energy $E_{vac}=-0.38$~eV below the
 5$d$ La conduction band at $\Gamma$ point which is a
$t_{2g}$ state and is decoupled from the vacancy state. Therefore,
electrons will occupy the less energetic vacancy state instead of the
conduction band, implying an insulating ground state for $x\leq
3$. When more vacancies are added, more localized states are formed
within the band gap and the material will become metallic when
localized states start to substantially overlap with each other.  

In Table \ref{vacdist} we list the probability of the vacancy state at
different atomic sites in the lattice. As we can see, the state is
largely distributed in the six neighboring La sites with the
probability of 58\%. This clearly shows its localized nature.

The localized nature of the vacancy state in LaH$_{x}$ is  consistent with 
the temperature dependence of the d.c. resistivity data at room temperatures,
which has a temperature depedence consistent with variable range hopping
~\cite{Shinar}. Localized states are a prerequisite for variable range hopping.
In the next section we will discuss the electronic band width of the vacancy
states.


\subsection{Effective electron hopping integrals between two  vacancies}

In LaH$_3$, octahedral hydrogens are not fixed in position but able to move
around by diffusion. Electrons in the vacancy state, of course, are
strongly bound by the Kondo like effect as discussed before. 
Therefore, a  calculation of the
vacancy-vacancy interaction and the effective electron hoppings between the
two H$^{\rm V}_{\rm oct}$ is necessary to the electronic band of the vacancy
states. We estimate here the effective hopping integrals between two 
H$^{\rm V}_{\rm oct}$ sites.

The effective hopping integral $t_{v-v}$ between the two H$_{\rm oct}$ vacancy 
sites is given by 
$$ t_{v-v} =\langle\Psi_1 \left| H' \right| \Psi_2 \rangle, $$

\noindent where $H'$ is the inter-atomic Hamiltonian given by H in 
Eq.~(\ref{Ham}). $|\Psi_1\rangle $ and 
$|\Psi_2\rangle$ are the electron state of the two vacancy sites. 

\begin{figure}[htbp]
\vspace{0.5cm}
\centerline{\psfig{figure=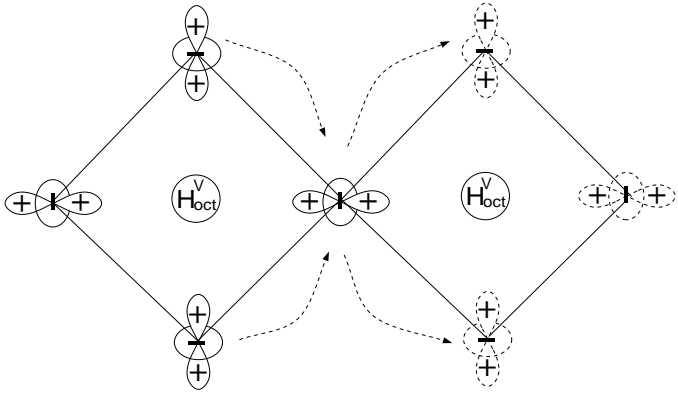,width=9cm}}
\vspace{0.5cm}
\caption{A 2-D illustration of vacancy state hopping. The dashed line with 
arrow represents the microscopic hopping process. Note that in 3-D, the
nearest neighbor vacancy states have two common La atoms and have a shorter
inter-vacancy distance.
\label{vachop}}
\end{figure}

Note that the vacancy states $|\Psi_{\alpha}\rangle$ are basically linear 
combinations of the nearest 6 La-$5d$-$e_g$ orbitals (Fig.~\ref{vachop}). 
For simplicity, we model the vacancy state by the local description as shown
in Fig.~\ref{vacancy}, and only the n.n. coupling between La-sites
in $H'$, and hence $V_{dd\sigma}$ and $V_{dd\pi}$, will be considered 
while contributions from further distant orbitals is negligible.
We apply the fitted values of $V_{dd\sigma}$ and $V_{dd\pi}$ from
section \ref{micmodel}. Table \ref{vachopint} lists the values of hopping 
integral of vacancy states separated by different distances.

It is interesting to notice that the value of the hopping integral oscillates 
when the distance increases. A more careful investigation tells us that the 
hopping of the vacancy state within the same cubic sublattice is larger in 
value and negative in sign.
In La hydrides H$_{\rm oct}$ sites forms a face-center cubic lattice, which
can also be divided into 4 interlocking simple cubic sublattices. Our
finding shows the hopping within the same sublattice is more probable than 
between different sublattices.

The precise nature of the transition in optical properties is not yet well 
understood.  Since the
vacancy states are highly localized, the vacancy concentration at the metal 
insulator transition of a random vacancy distribution is quite high.
 Hence the
mobile carrier density is high. This explains
why the metallic phase is a good metal with high reflectivity.

One aspect of the hydride system is unique, namely the existence of an
order-disorder transition
among the vacancies which enables one to separate the effects of disorder and
Coulomb interaction in the electronic band of vacancy states.  Usually donors
and acceptors are fixed in doped semiconductors so that this is impossible.
The disordered vacancy phase is insulating with variable range hopping among
the localized vacancy states.  Interestingly Shinar $\it{et} \
\it{al.}$~\cite{Shinar}
reported a transition to an ordered vacancy phase at temperatures around 
250$^{\rm o}$K
for concentrations $x \sim 2.8 -2.9$ and metallic d.c. conductivity in the
ordered
phase.  This shows that Anderson localization is responsible for the
insulating character of the disordered phase. However in the ordered phase
there must be a Mott transition to an insulator as the vacancy density is
lowered.  It would be very interesting to examine this region and look for the
Mott transition in an ordered vacancy sublattice.


\section{PHYSICAL PROPERTIES OF INSULATING LaH$_{3-\delta}$}

In order to study the electronic structure of metal hydrides, various
measurements, such as photoelectron and optical spectroscopy, have been 
performed for metal hydrides with different hydrogen concentration. From
these experiments \cite{Peterman} interband excitations can be measured and 
compared to
theoretical band structure. Their data can be used to verify the
accuracy of theoretical prediction for the electronic structure. Most
theoretical investigations  have been done on the dihydrides and 
trihydrides, while hydrides with intermediate hydrogen concentration are
difficult to study. In this section we will discuss some physical
properties predicted from our theory in connection with the existing or
further experiments.


\subsection{Magnetic properties of LaH$_{3-\delta}$}

The vacancy state is a localized object of spin-1/2 as discussed in section 
\ref{secvac}. Near the trihydride phase, LaH$_{3-\delta}$ is an insulator for
small $\delta$. The vacancies have a tendency to be singly occupied because
two electrons on the same vacancy site repel each other. The magnetic
susceptibility is expected to follow a Curie law with a linear
inverse temperature dependence. 


\subsection{Density of states}

\begin{figure}[htbp]
\vspace{0.5cm}
\centerline{\psfig{figure=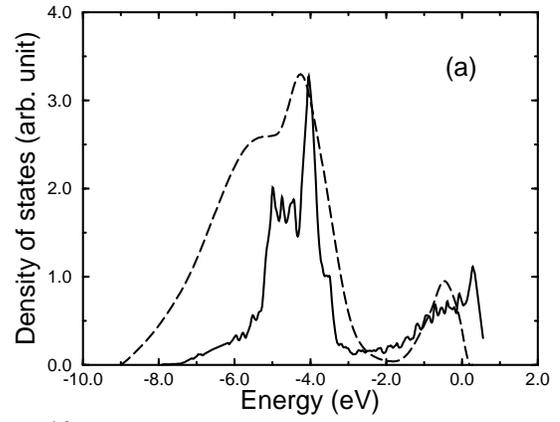,height=5.5cm}}
\centerline{\psfig{figure=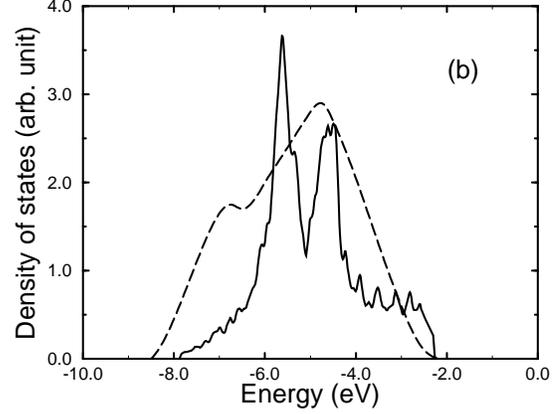,height=5.5cm}}
\vspace{0.5cm}
\caption{Density of states for (a) LaH$_2$ and (b) LaH$_3$. Solid curves are 
our calculated results and dashed curves are the experimental data obtained 
from Peterman $et$ $al.$ \protect \cite{Peterman}.
\label{XPS}}
\end{figure}

Peterman $et$ $al.$ \cite{Peterman} studied the composition-dependent 
electronic structure
of LaH$_x$, $1.9\leq x \leq 2.9$, using photoelectron spectroscopy with 
synchrotron radiation ($10$~eV $\leq h\nu  \leq 50$~eV). Fujimori $et$ $al.$
\cite{Fujimori} on the other hand studied the electronic structure of yttrium 
hydride by x-ray photoemission spectroscopy. 
Here we compare our calculated densities of states with the experiment on 
lathanum hydrides, as shown in Fig.~\ref{XPS}. 

We performed a finite size calculation with lattice size
of $15\times15\times15 a_0^3$ based on the model Hamiltonian (\ref{Ham}). 
Numbers of states in the band structure of LaH$_3$ and LaH$_2$ 
are counted as a function of energy. The discrete densities of states were 
smoothed out by replacing delta peaks with Lorentzian distribution curve of 
width $\Gamma=0.3$~eV. Our results for LaH$_2$ and LaH$_3$ are similar to the 
previous theoretical calculations by Gupta and Burger \cite{Peterman,Gupta}. 
In Fig.~\ref{XPS}, the pronounced two-peak structure is associated with the 
flat region of hydrogen bands near symmetry points X, L, K (refer to the
band structure Fig.~\ref{gut1} and \ref{gut2}) while the small bump at the 
zero energy in LaH$_2$ is associated with the La band near symmetry points 
W and K. Experiments showed that the small bump shrinks as hydrogen 
concentration increases and eventually disappears when the sample approaches 
the trihydride. However Gupta and Burger expected a metallic state for LaH$_3$
from their LDA calculation which ignored electron correlation. In our 
calculations, the density of states goes to zero at $-2.2$~eV, indicating an 
insulating behavior for LaH$_3$. However, the calculated widths for LaH$_2$ 
and LaH$_3$ are considerably smaller than the experiment results, a 
discrepancy similar to that in the LDA . The experimental data also 
shows a shift of the lowest energy peak when the concentration increases. The 
positions of calculated peaks do not significantly depend on the concentration
and does not agree to the experiment.
The origin of these discrepancies is not clear. Since our theory
uses some parameters extracted from LDA, similar features except for the band 
gap may be expected from both theories.


\subsection{Optical Conductivity}

The optical absorptivity spectra for samples of LaH$_x$ and NdH$_x$ were also
measured by Peterman $et$ $al.$ \cite{Peterman} at 4.2K at near-normal 
incidence. 
From the absorption data, they deduced the real and imaginary part of the 
dielectric constant, and hence the optical conductivity  
 by using Kramers-Kronig analysis. They found a relatively 
broad feature for LaH$_{2.87}$. However, their samples were
polycrystals and the use of Kramers-Kronig analysis might enlarge the
data uncertainties. On the other hand, a more recent and accurate measurement 
reported by Griessen $et$ $al.$ \cite{Griessen1} sheds new light on the 
optical conductivity curve. The optical transmission spectra of the insulating
phase YH$_{3-\delta}$ were measured as functions of the photon frequency 
$\hbar \omega$ and of hydrogen vacancy concentration $\delta$. The effect of
the vacancy appears to reduce the overall transmission spectra quite evenly
between $\hbar \omega=0.5$~eV and 2~eV. In our theory, the $\delta$-dependent 
conductivity $\sigma(\omega)$ at $\hbar\omega<2.8$~eV mainly determined by the
optical transition from the vacancy state to the conduction bounds. The 
transition energy from the valence bands to the vacancy state is larger because
of the larger energy difference between the two states and because of the 
Coulomb repulsion of the doubly occupied electron states on the same vacancy
site. Since the vacancy state is highly localized, the transition matrix 
largely depends on the density of states of the conduction bands, which is
expected to lack pronounced feature.

\begin{figure}[htbp]
\vspace{0.5cm}
\centerline{\psfig{figure=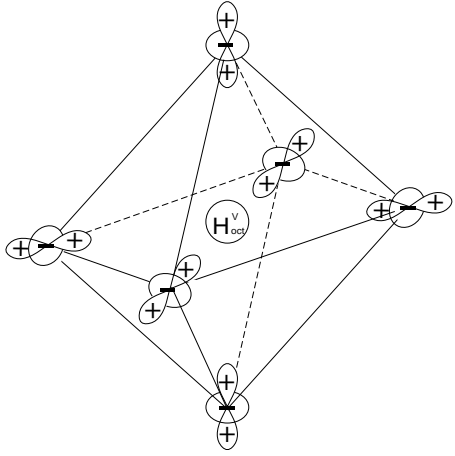,height=6cm}}
\vspace{0.5cm}
\caption{Diagrammatic illustration of the proposed H$_{\rm oct}$ vacancy 
state in LaH$_3$. The center circle represents a H vacancy, forming an 
$n$-type impurity center. The surrounding orbits represent phases for the
local $s$-like octahedral La-5$d$-$e_g$. The vacancy state has spin-1/2.
\label{vacancy}}
\end{figure}

Within the dipole approximation, the real part of conductivity contributed from
the vacancies is given by 

\begin{equation}
\sigma\left(\omega\right)=const\times\omega\sum_{n}\left|\langle 
n\left|x\right|vac\rangle\right|^2  \delta\left(
E_n-E_{vac}-\hbar\omega\right)
\label{omega(w)}
\end{equation} 

\noindent where $|vac\rangle$ is the vacancy state. The sum is over all 
conduction states $|n\rangle$. We shall use a local model to describe 
$|vac\rangle$, as shown in Fig.~\ref{vacancy},

\begin{equation}
|vac\rangle=\sum_{{\bf R}_{La},\alpha} a_\alpha ({\bf R}_{La})|{\bf R}_{La}
,d_\alpha\rangle, 
\end{equation}

\noindent where ${\bf R}_{La}$ denotes the position of the 6 La atoms 
and $d_\alpha$ the $5d$-$e_g$ orbitals, and $a_{\alpha}({\bf R}_{La})$ are
the wavefunction amplitudes. The conduction bands are mainly 
$5d$-La character mixed with $1s$-H orbitals. The La intraband transition 
is a difficult case to calculate with the unknown transition 
matrices involved. For simplicity and for the purpose of illustration, here
we consider the part of the optical transition to the conduction bands with
the H$_{\rm tet}$ character, and assume the transition matrix is non-zero only
between the La-$5d$ and to its neighboring H$_{\rm tet}$ orbitals. 

\begin{equation}
|n\rangle=\sum_{{\bf R}_{t}} b_n({\bf R}_{t})|{\bf R}_{t},s\rangle, 
\end{equation}

Let $|{\bf R}_{t},s\rangle$ be the H$_{\rm tet}$ $1s$ state at site 
${\bf R}_{t}$, and $b_n({\bf R}_{t})=\langle{\bf R}_t,s|n\rangle$ the amplitude
of the state $|{\bf R}_{t},s\rangle$ in $|n\rangle$, we then have

\begin{equation}
\langle vac\left|x\right|n\rangle =
\sum_{{\bf R}_{La},{\bf R}_t,\alpha} a^*_\alpha({\bf R}_{La})b_n({\bf R}_t)
\langle {\bf R}_{La},d_\alpha\left|x\right|{\bf R}_t,s\rangle,
\end{equation}

\noindent where the sum over ${\bf R}_t$ runs all the neighboring H$_{\rm tet}$
of the six La sites. We calculate  The $a^*_\alpha({\bf R}_{La})$ and 
$b_n({\bf R}_t)$ by numerically solving a finite size cell 
($15\times 15\times 15 a_0^3$) of LaH$_3$ with a single H$_{\rm oct}$ vacancy.
The orbital overlap
$\langle d_{\alpha}|x|s\rangle \propto \langle d_\alpha|p_x\rangle$ is a 
linear combination of the coupling parameters $V_{pd\sigma}$ and $V_{pd\pi}$ 
and can be found in Harrison's book \cite{Harrison}. It is estimated the ratio 
$V_{pd\sigma}/V_{pd\pi}\simeq -2.17$. The optical conductivity 
$\sigma (\omega)$ can thus be calculated up to an overall constant. In the 
calculations, we use a rigid band approximation, where the conduction band 
spectra is unaffected by the vacancy. This approximation is 
justified by comparing the result to the actual conduction band in the presence
of the vacancy from the finite size calculation. Only minute deviation is 
observed in small size calculation. 

\begin{figure}[htbp]
\vspace{0.5cm}
\centerline{\psfig{figure=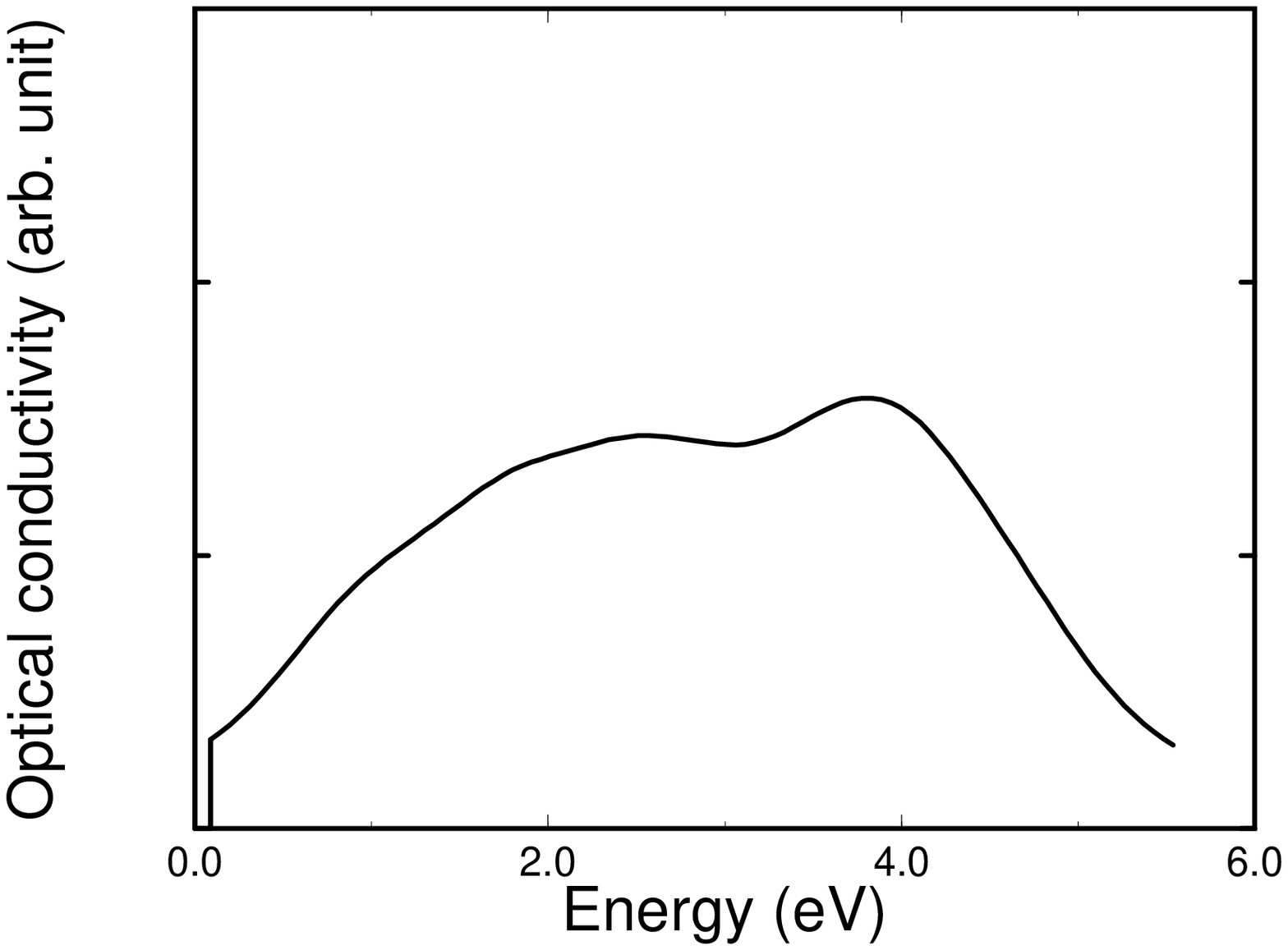,height=5.5cm}}
\vspace{0.5cm}
\caption{Optical conductivity due to excitation of electrons from vacancy 
state to conduction band of LaH$_3$.
\label{optcond}}
\end{figure}

The calculated result (Lorentzian width 0.5~eV) is showed in the 
Fig.~\ref{optcond} which gives a fairly featureless curve. Except the small 
peak around 4 eV, the curve does not show any pronounced features. This is
qualitatively consistent with the experiments \cite{Peterman,Griessen}.

There is a threshold $\sim E_c$ in $\sigma(\omega)$, which is the energy 
cost for the vacancy state to transfer to the conduction band. In our 
calculation, $E_c \simeq 0.1$~eV. This small value is due to the inaccurate
tight-binding fitting to the LDA, resulting in lower energy at $X$-point in 
LaH$_3$. If we subtract this, $E_c \simeq 0.38$~eV. In our 
estimate, a charged Coulomb attraction has not been added, so that the actual 
threshold is larger.

It is interesting to compare the optical properties of the hydrides vacancy 
state with those of the usual doped
semiconductors studied by Thomas $et$ $al$ \cite{Thomas}. First we note an
important difference between the two systems. In a conventional semiconductor,
the hydrogen-like $1s$ groundstate has an electronic dipole active transition
to the bound $2p$ state, leading to optical absorption peaks at about 30~meV in
Si. In the metal hydrides, the ground state of a vacancy is a highly 
localized $s$-wave-like state (see Fig.~\ref{vacancy}).

However, the  $p$-wave-like state is expected to be only weakly bound with much
larger spatial distribution.  This is because of the absence of short-range 
attraction as in the s-wave state.  The long range Coulomb attraction 
between the electron and the vacancy induces  a bound p-wave 
state with a larger radius
similar to the hydrogen-like ones.  Therefore we expect
only a  weak dipole transition intensity  from the very localized s-wave
state (mainly induced by the short range bonding force) to the p-wave state.
The absence of the short-range force in the p-wave vacancy state 
 can be explained  as follows. 
Consider a local $p$-wave-like state 
of La-$5d$ orbits, similar to that in Fig.~\ref{vacancy} except for the 
symmetry that is changed from $s$- to $p$-wave-like. By symmetry, the 
$p$-wave-like state is decoupled to the $1s$ electron state of the center 
hydrogen atom. Note that this
property of the vacancy state is markedly different from that of the negative
ion vacancy in an alkali halide crystal. The latter is called a color center 
that has one excess electron bound at the vacancy \cite{Kittel}. The potential
created by the vacancy in alkali halides is a long-range Coulomb force, and the
color center absorbs visible light in a dipole transition to a bound excited
state of the center. Our theory for the vacancies in hydrides predicts 
much weakness  of such transitions.


\section{Conclusion}

In this paper, we have examined the importance of the electron correlation in
negatively charged ions (H$^-$ ions), and identify the metal hydrides to be a
strongly correlated system. We develop a many-body theory to describe
the H$^-$ ion lattice in hydrides. We use lanthanum hydrides as a prototype
for these compounds and show that LaH$_2$ is a metal, and LaH$_3$ is a band
insulator where the electron correlation plays a crucial role.

The electronic structure for metal hydrides is described by the large $U$-limit
Anderson lattice model. The parameters for the hydrides are determined by a
combination of  method including the microscopic calculations for the
electron hopping integrals between hydrogen sites and parameter value
obtained from LDA methods for La/H hybridization. The 
Anderson lattice model is then solved using a many-body technique 
$-$ Gutzwiller method. The elementary entity when a H atom is removed
from the insulating LaH$_3$ is the
octahedral hydrogen vacancy, which is highly localized electronically,
due to the strong  
hybridization between the H-1$s$ and its neighboring La-5$d$-$e_g$ orbitals. 
This explains why the metal-insulator transition can occur at a large vacancy 
concentration $\sim 25 \%$. 

Our theory for the localized vacancies in hydrides
predicts a much weak optical transition to the hydrogen-like-$2p$ 
state in comparison with observed transitions in conventional semiconductors.

\section{Acknowledgment}

We thank Prof. R. Griessen, O. Gunnarsson for many useful discussions.
This work was in part supported by DOE grant DE/FG03-98ER45687, and by
Russian Foundation for Basic Research grant RFFI-98-02-17275. 
KKN, FCZ and VIA would like to thank the Zentra for the Studies of ETH for
support during their visit to ETH where this project was started. 
FCZ and KKN
wish to acknowledge the URC support from University of Cincinnati.

\vbox{
\begin{table}[b]
\vspace{0.5cm}
\centering
\begin{tabular}{lccc}
$d(a.u.)$ & 4 & 4.5 & 6 \\
\hline 
$E_1$(a.u.) & -0.25746 & -0.22687 & -0.16789 \\
$a_1$       & 0.44324  & 0.39654  & 0.27863  \\
$a_2$       & 0.08514  & 0.05775  & 0.01804  \\
$a_3$       & 0.02785  & 0.01376  & 0.00146  \\
$b_1$       & -0.14039 & -0.11154 & -0.05866 \\
$b_2$       & -0.04145 & -0.02566 & -0.00748 \\
$b_3$       & -0.01609 & -0.00731 & -0.00123 \\
$t$         & -0.03611 & -0.02857 & -0.01346 \\
\end{tabular}
\vspace{0.5cm}
\caption{Values of integrals in Eq.~(\protect\ref{integrals}) and the
hopping integrals $t(d)$ for different inter-proton distances $d$. All 
distances and energies are in atomic unit.
\label{int}} 
\end{table}
}

\vbox{
\begin{table}[b]
\centering
\vspace{0.5cm}
\begin{tabular}{cc}
Cubic size ($a_0$) & Crystal field reduction (eV) \\
\hline
1 & 0.084674\\
2 & 0.145003\\ 
3 & 0.212051\\
4 & 0.223285\\
5 & 0.226712\\
6 & 0.230330\\
\end{tabular}
\vspace{0.5cm}
\caption{Change of $t_1$, the electron hopping integral between the n.n. 
tetrahedral H sites,  due to crystal field effect of ions within a cube
cell of various size.  The change converges as the size of the cubic cell 
increases. $a_o$: lattice constant in LaH$_3$.
\label{crysfield}} 
\end{table}
}

\vbox{
\begin{table}[b]
\vspace{0.5cm}
\centering
\begin{tabular}{ll}
Parameters & fitting values\\
\hline 
atomic energy of La 5$d$ $t_{2g}$        & $\epsilon_d=1.6$\\
atomic energy of La 5$d$ $e_g$           & $\epsilon'_d=1.4$\\
atomic energy of H$_{\rm tet}$      & $\epsilon_t=-3.2$\\
atomic energy of H$_{\rm oct}$      & $\epsilon_o=-2.6$\\
n.n. La-La $\pi$-bonding            & $V_{dd \pi}=0.6$\\
n.n. La-La $\sigma$-bonding         & $V_{dd \sigma}=-1.2$\\
n.n.n. La-La $\pi$-bonding       & $V'_{dd \pi}=-0.1$\\
n.n.n. La-La $\sigma$-bonding    & $V'_{dd \sigma}=-0.2$\\
n.n. La-H$_{\rm oct}$ $\sigma$-bonding   & $V^{La-o}_{sd \sigma}=-1.2$\\
nearest La-H$_{\rm tet}$ $\sigma$-bonding   & $V^{La-t}_{sd \sigma}=-1.3$\\
nearest H$_{\rm tet}$-H$_{\rm tet}$ $\sigma$-bonding & 
$t_{t-t}=-0.4$\\
nearest H$_{\rm oct}$-H$_{\rm tet}$ $\sigma$-bonding & 
$t_{t-o}=-0.79$\\
\end{tabular}
\vspace{0.5cm}
\caption{Tight binding parameters extracted from the LDA results of 
Fig.~\protect\ref{vladband} for LaH$_3$ and LaH$_2$. All the energies are in 
units of eV. The band structure of LaH$_3$ from the tight binding model of 
these parameters is plotted in Fig.~\protect\ref{fitband}.
\label{parameter}}
\end{table}
}

\vbox{
\begin{table}[b]
\vspace{0.5cm}
\centering
\begin{tabular}{ll}
Parameters & Renormalized values\\
\hline 
nearest La-H$_{\rm oct}$ $\sigma$-bonding   & $V^{La-o}_{sd \sigma}=-1.004$\\
nearest La-H$_{\rm tet}$ $\sigma$-bonding   & $V^{La-t}_{sd \sigma}=-1.146$\\
nearest H$_{\rm tet}$-H$_{\rm tet}$ $\sigma$-bonding 
& $V^{t-t}_{ss \sigma}=-0.228$\\
nearest H$_{\rm oct}$-H$_{\rm tet}$ $\sigma$-bonding 
& $V^{o-t}_{ss \sigma}=-0.309$\\
\end{tabular}
\vspace{0.5cm}
\caption{Renormalized parameters of Gutzwiller method for LaH$_3$.
\label{gutparam}}
\end{table}
}

\vbox{
\begin{table}[b]
\vspace{0.5cm}
\centering
\begin{tabular}{cccc}
type of atom  & position $(a_0)$ & distance $(a_0)$ & weighting\\
\hline
H$_{\rm{tet}}$ & $(\frac{1}{4},\frac{1}{4},\frac{1}{4})$ & 0.4330 & 0.0012\\ 
H$_{\rm{tet}}$ & $(\frac{3}{4},\frac{1}{4},\frac{1}{4})$ & 0.8292 & 0.0365\\
H$_{\rm{tet}}$ & $(\frac{3}{4},\frac{3}{4},\frac{1}{4})$ & 1.0897 & 0.0002\\
H$_{\rm{tet}}$ & $(\frac{3}{4},\frac{3}{4},\frac{3}{4})$ & 1.2990 & 0.0134\\ 
H$_{\rm{tet}}$ & $(\frac{5}{4},\frac{1}{4},\frac{1}{4})$ & 1.2990 & 0.0087\\
H$_{\rm{tet}}$ & $(\frac{5}{4},\frac{3}{4},\frac{1}{4})$ & 1.4790 & 0.0012\\ 
H$_{\rm{oct}}$ & $(\frac{1}{2},\frac{1}{2},0)$           & 0.7071 & 0.1426\\
H$_{\rm{oct}}$ & $(1,0,0)$				 & 1.0000 & 0.0111\\
H$_{\rm{oct}}$ & $(1,\frac{1}{2},\frac{1}{2})$	         & 1.2247 & 0.0001\\
H$_{\rm{oct}}$ & $(1,1,0)$   		                 & 1.4142 & 0.0004\\
La             & $(\frac{1}{2},0,0)$                     & 0.5000 & 0.5790\\
La             & $(\frac{1}{2},\frac{1}{2},\frac{1}{2})$ & 0.8660 & 0.0983\\
La             & $(\frac{1}{2},1,0)$  	                 & 1.1180 & 0.0499\\
La             & $(\frac{3}{2},0,0)$                     & 1.5000 & 0.0269\\
\end{tabular}
\vspace{0.5cm}
\caption{Weighting factors of the vacancy state at different atomic sites. 
The third column shows the total weighting of atomic 
sites which share the same distance from the vacancy site (origin). 
Only one position vector is shown in the second column for each 
distance while the others can be deduced from symmetry.
\label{vacdist}}
\end{table}
}

\vbox{
\begin{table}[b]
\vspace{0.5cm}
\centering
\begin{tabular}{cc}
 Vacancies separation ($a_o$)& Hopping integral (eV) \\
\hline
$\sqrt{2}$/2 (different s.l.) & 0.475 \\
1 (same s.l.) & -0.7     \\
$\sqrt{3/2}$ (different s.l.) & 0.125  \\
$\sqrt{2}$  (same s.l.) & -0.175  \\
$\sqrt{10}/2$ (different s.l.) & 0.0625 \\
\end{tabular}
\vspace{0.5cm}
\caption{Calculated hopping integrals of vacancy state for various vacancies
separation. Inside the parentheses are the indication of the two vacancies 
belonging to the different or the same sublattices (s.l.) of the fcc.
\label{vachopint}}
\end{table}
}

\begin{figure}[htbp]
\vspace{0.5cm}
\centerline{\psfig{figure=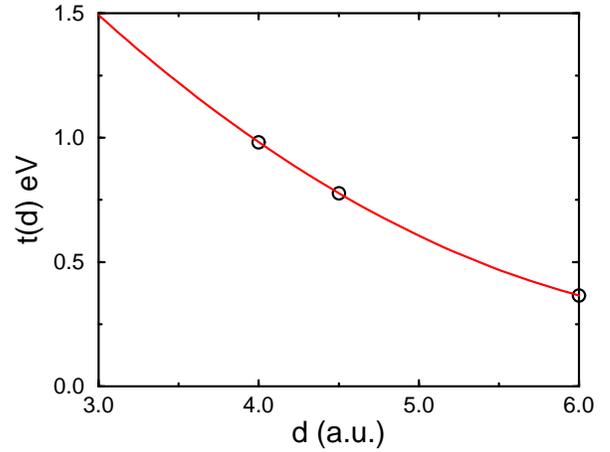,height=6cm}}
\vspace{0.5cm}
\caption{Electron hopping integral $t(d)$ as a function of the distance $d$ 
between two
 H ions. The hydrogen hopping integral decreases rapidly with increasing
distance between the two H-ions. The equilibrium separation between the two
nearest H$_{\rm tet}$'s, and between neighboring H$_{\rm tet}$ and 
H$_{\rm oct}$ are 5.29 a.u. and 4.58 a.u., respectively. 
The solid line shown here is fitted to the three data points.
\label{hop} }
\end{figure}

\begin{references}
\bibitem{Kondo} The bound H$^-$-ion is a spin singlet electron state.
We consider here the spin triplet electron state to be essentially non-bound.
\end{references}
\end{document}